\documentclass[superscriptaddress,groupedaddress,nofootnoteinbib,11pt]{article}
\pdfoutput=1 
\usepackage{jheppub}

\makeatletter
\gdef\@fpheader{}
\makeatother

\usepackage{multicol}
\usepackage{amsmath}
\usepackage{amsfonts}
\usepackage{amssymb}

\usepackage{blindtext}
\usepackage{lipsum}
\usepackage{afterpage}
\usepackage{sectsty}
\usepackage{graphicx}
\usepackage{xcolor}
\usepackage[dvipsnames]{xcolor}
\usepackage{float}
\usepackage{slashed}
\usepackage{lscape}
\usepackage{graphicx} 
\usepackage{url}
\usepackage{subcaption}

\usepackage[normalem]{ulem} 

\usepackage{tikz}
\usepackage{tikz-feynman,contour}
\usepackage{subcaption}

\graphicspath{{./sn-higgs-plots/}}
\numberwithin{equation}{section}

\title{Improved supernova bounds on CP-even scalars: cooling and decay constraints}
\author[a]{Melissa Joseph}
\author[a]{Samuel Liebersbach}
\author[a]{Anirudhan A. Madathil}
\author[a]{Gustavo Marques-Tavares}
\date{March 2025}

\affiliation[a]{Department of Physics and Astronomy, University of Utah, Salt Lake City, UT, 84112, USA}

\abstract{
Supernovae provide among the most powerful probes of weakly-coupled new particles 
in the MeV mass range, where laboratory experiments lose sensitivity. 
In this work, we derive improved supernova constraints on CP-even scalars mixing 
with the Higgs boson, combining an updated production rate calculation, which 
improves the cooling bound by more than an order of magnitude, with new 
decay-based constraints from the galactic 511~keV positron flux and energy 
deposition in low-energy Type~II-P supernovae. Together, these constraints probe 
mixing angles as small as $\sin\theta \sim 10^{-9}$, more than five orders of 
magnitude below existing collider bounds. We also extend our analysis to a 
hadrophilic scalar model, constraining Yukawa couplings down to 
$y_N \sim 10^{-10}$. Our results demonstrate that the combination of astrophysical 
and collider probes covers over nine orders of magnitude in coupling for these 
classes of models, probing a large region of parameter space motivated by dark 
matter considerations.
}

\begin{document}
\newcommand{\mbf}[1]{\mathbf{#1}}
\newcommand{\red}[1]{\textcolor{red}{#1}}
\newcommand{\blue}[1]{\textcolor{blue}{#1}}

\newcommand{\beq}{\begin{equation}}
\newcommand{\eeq}{\end{equation}}
\newcommand{\MJ}[1]{\textcolor{orange} {\textsf{[MJ: #1]}}}
\newcommand{\AAM}[1]{\textcolor{blue} {\textsf{[AAM: #1]}}}
\newcommand{\Sl}[1]{\textcolor{ForestGreen} {\textsf{[SL: #1]}}}

\maketitle

\section{Introduction}

The absence of new physics signals at colliders has renewed interest in light, weakly-coupled dark sectors as a viable frontier for beyond-the-Standard-Model (BSM) physics~\cite{Alexander:2016aln,Batell:2022dpx,Gori:2022vri,Antel:2023hkf,Alimena:2025kjv}. A systematic framework for studying such sectors is to classify the lowest-dimension gauge-invariant operators connecting Standard Model (SM) and gauge-singlet fields, known as portal operators. For a CP-even scalar $S$, the lowest-dimensional allowed operator is the dimension-3 coupling $S H^\dagger H$, known as the Higgs portal~\cite{Krnjaic:2015mbs}.%
\footnote{A dimension-4 operator preserving a $\mathbb{Z}_2$ symmetry $S \rightarrow -S$, namely $\lambda S^2 |H|^2$, is also often referred to as the Higgs portal in the literature. The two scenarios share similar phenomenology when the scalar acquires a vev, which is the scenario we will explore in this work.}
After electroweak symmetry breaking, this operator induces mixing between $S$ and the Higgs boson, generating couplings to SM fermions that are proportional to their masses and suppressed by the Higgs vacuum expectation value $v$,
\begin{equation}
    \mathcal{L} \supset -\sin\theta \frac{m_f}{v} S \bar{f} f,
\end{equation}
where $\theta$ is the Higgs--scalar mixing angle. Higgs-portal scalars of this type arise across a broad class of models and have been widely studied in the context of dark matter production, where the portal coupling governs the connections between the SM and the hidden sector~\cite{Krnjaic:2015mbs,Knapen:2017xzo}. In the small coupling regime, laboratory and collider probes lack sensitivity, making astrophysical environments --- and supernovae (SN) in particular --- among the most powerful available constraints~\cite{Raffelt:1996wa}.

A new CP-even scalar has been studied in various contexts. It can mediate interactions between the visible sector and a dark sector~\cite{Pospelov:2007mp,Piazza:2010ye,Pospelov:2011yp,Schmidt-Hoberg:2013hba,Kouvaris:2014uoa,Krnjaic:2015mbs,Kainulainen:2015sva,Bell:2016ekl,Matsumoto:2018acr,Hooper:2025iii}
 or extra dimensions~\cite{Diener:2013xpa}, or play the role of dark matter~\cite{Silveira:1985rk,McDonald:1993ex,Burgess:2000yq,Bird:2004ts,Bird:2006jd,He:2007tt,Barger:2007im,Ponton:2008zv,Cline:2013gha,GAMBIT:2017gge}. It may also allow electro-weak phase transitions~\cite{Ham:2004cf,Profumo:2007wc,Espinosa:2011ax} and control radiative corrections to the Higgs mass~\cite{Bazzocchi:2012de}. There have been numerous efforts to probe and constrain the parameter space of the CP-even scalar~\cite{ Winkler:2018qyg,Gori:2022vri,Batell:2022dpx}. High-energy astrophysical phenomena, such as core-collapse SN, can probe the parameter space of the MeV-scale scalar with small $\sin \theta$, which otherwise cannot be probed in ground-based experiments~\cite{Krnjaic:2015mbs,Dev:2020eam,Balaji:2022noj,Hardy:2024gwy}. The core of the proto-neutron star (PNS) formed during a core-collapse SN is extremely dense and hot, making it a conducive environment for creating MeV-range scalar particles even if they are very weakly coupled. In this paper, we derive improved SN bounds on CP-even scalars, combining an updated and improved production calculation with new constraints from scalar decays to visible particles.

BSM particles produced inside the PNS during the SN explosion could alter the PNS's cooling dynamics, resulting in a shorter neutrino burst which would contradict the SN1987a neutrino observations. This is called the Raffelt criterion or SN cooling bound~\cite{Raffelt:1996wa}. SN cooling bounds in the context of CP-even scalars have been previously explored~\cite{Dev:2020eam,Hardy:2024gwy,Krnjaic:2015mbs,Balaji:2022noj}. In this work, we improve the existing cooling bounds on scalars by improving the production rate calculation of scalars in the one-pion exchange approximation. The improvement in the production rate stems from performing the non-relativistic expansion of the nucleon propagator in a manner consistent with the relevant hierarchy of scales in the SN environment. We find that contributions from higher order terms in the nuclei momenta expansion increase the production rate by an order of magnitude for scalars lighter than 100 MeV, leading to significant changes in the cooling bound.  We also correct the interaction of the scalar with the pion in Ref.~\cite{Dev:2020eam, Balaji:2022noj}, which increases the contribution from diagrams in which the scalar is emitted from an internal pion propagator.

Apart from the cooling bound, the decay of scalars produced inside the SN core back to SM particles can have distinct observational consequences. Other observational SN bounds have been studied in the context of dark photons~\cite{DeRocco:2019njg,Sung:2019xie,Calore:2021lih}, axions/axion like particles~\cite{Payez:2014xsa,Jaeckel:2017tud,Calore:2020tjw,Calore:2021klc,Caputo:2022mah,Hoof:2022xbe,Calore:2023srn,Muller:2023pip,Buckley:2024ldr,Ferreira:2025qui,Chauhan:2025xqr}, and sterile neutrinos~\cite{Calore:2021lih, Chauhan:2023sci, Carenza:2023old,Chauhan:2025mnn}. In this work, we extend this to the CP-even scalar. We explore the parameter space further by requiring that the decay of scalars should not lead to an overabundance of galactic positrons. More constraints can be explored in the context of low-energy SN (LE-SN) by requiring the energy deposited by scalars back into the progenitor mantle to not exceed the observed explosion energy of low-energy Type II-P SN. The positron and LE-SN bounds further constrain the parameter space of the CP-even scalar.

Another class of CP-even scalars is the hadrophilic scalar $S_{N}$, in which the scalar has negligible interactions with leptons. Hadrophilic scalars could mediate the interaction between dark matter and the SM~\cite{Knapen:2017xzo, Benato:2018ijc,Batell:2018fqo,Alvey:2019zaa,Batell:2021snh,Arguelles:2022fqq,Elor:2021swj,Cox:2024rew,Gori:2025jzu}. Thus, these scalars could produce signals in nuclear recoil direct-detection experiments. Such a scalar could also be produced in SN through nucleon-nucleon bremsstrahlung, but their decay lengths are generally much longer. We generalize our previous calculations to determine the SN cooling bounds on a simple hadrophilic scalar model in which the coupling originates from the dimension 5 operator $S_N G^2$.

The rest of this paper is structured as follows. In Section~\ref{sec:production}, we discuss the scalar production mechanism and the calculation of the scalar luminosity. Following this, we discuss the decay channels and absorption mechanisms of the scalar in Section~\ref{sec:abs_and_decay}. We then discuss all the relevant bounds explored in Sections~\ref{sec:rafbound}-\ref{sec:low_energy}. In Section~\ref{sec:nucleophilic}, we discuss the bounds on the hadrophilic scalar, and in Section~\ref{sec:conclusion}, we present our conclusions.

\section{Scalar Production}\label{sec:production}

The dominant production channel for scalars, $S$, in the SN core is nucleon-nucleon brems\-strahlung, $NN\rightarrow NNS$,
due to the large density of nuclei in the core and the relatively small induced couplings to electrons and photons. The calculation of the nucleon bremsstrahlung rate in a proto-neutron star is challenging both due to the typical center of mass energy scale being intermediate between the regime of validity of chiral perturbation theory and perturbative QCD, and due to the interacting nature of the fluid. For simplicity, we will compute this rate using the one-pion exchange (OPE) approximation in vacuum, and argue why, in this scenario, this is more appropriate than utilizing the soft-radiation approximation. Our results differ significantly from recent works utilizing the same approach~\cite{Dev:2020eam,Balaji:2022noj}, most noticeably in the low mass regime due to more consistent approximations in expanding the intermediate fermion propagators. It is worth emphasizing that it is known that the OPE approximation is not ideal for the conditions we explore, and in other scenarios where an alternative approach was available often differs by up to a factor of 2-3 from the other calculations. In what follows we describe the most important aspects of the computation, and the changes compared to previous work, for the full calculation see Appendix~\ref{appendix:amplitude}.

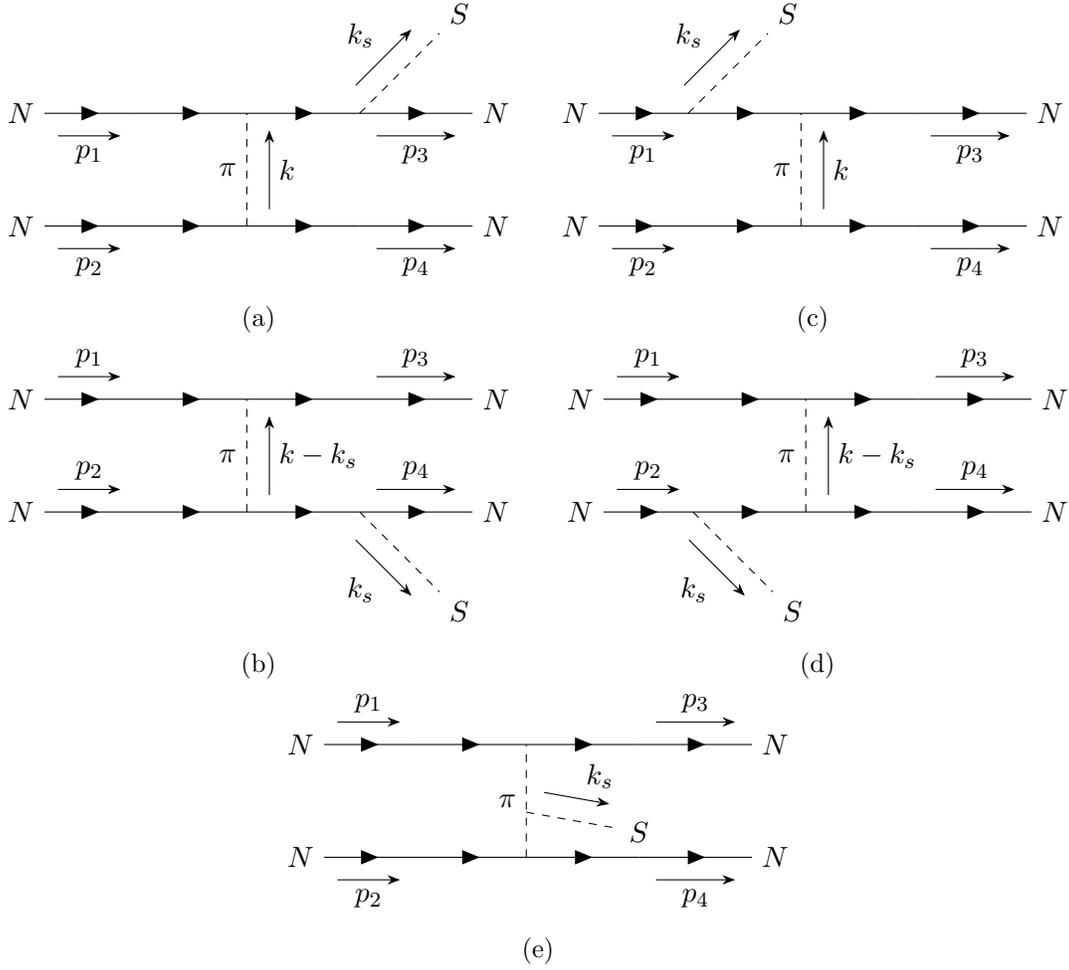
\begin{figure}[t!]
\captionsetup[subfigure]{labelformat=empty}
    \centering
    \subfloat[(a)]{\begin{tikzpicture}
    \begin{feynman}
        \vertex (a){\(N\)};
        \vertex [right= of a](b);
        \vertex [right= of b](c);
        \vertex [right= of c](d);
        \vertex [right= of d](e){\(N\)};
        \vertex [below= of a](j){\(N\)};
        \vertex [right= of j](i);
        \vertex [right= of i](f);
        \vertex [right= of f](g);
        \vertex [right= of g](h){\(N\)};
        \vertex [above right= of d] (p){\(S\)};

        \diagram*{
        (a)--[fermion,momentum'=\(p_{1}\)](b)--[fermion](c)--[fermion](d)[fermion]--[fermion,momentum'=\(p_{3}\)](e),
        (j)--[fermion,momentum'=\(p_{2}\)](i)--[fermion](f)--[fermion](g)--[fermion,momentum'=\(p_{4}\)](h),
        (f)--[scalar,momentum'=\(k\),edge label=\(\pi\)](c),
        (d)--[scalar,momentum=\(k_{s}\)](p),
        };
    \end{feynman}
\end{tikzpicture}}\quad
    \subfloat[(c)]{\begin{tikzpicture}
    \begin{feynman}
        \vertex (a){\(N\)};
        \vertex [right= of a](b);
        \vertex [right= of b](c);
        \vertex [right= of c](d);
        \vertex [right= of d](e){\(N\)};
        \vertex [below= of a](j){\(N\)};
        \vertex [right= of j](i);
        \vertex [right= of i](f);
        \vertex [right= of f](g);
        \vertex [right= of g](h){\(N\)};
        \vertex [above right= of b] (p){\(S\)};

        \diagram*{
        (a)--[fermion,momentum'=\(p_{1}\)](b)--[fermion](c)--[fermion](d)[fermion]--[fermion,momentum'=\(p_{3}\)](e),
        (j)--[fermion,momentum'=\(p_{2}\)](i)--[fermion](f)--[fermion](g)--[fermion,momentum'=\(p_{4}\)](h),
        (f)--[scalar,momentum'=\(k\),edge label=\(\pi\)](c),
        (b)--[scalar,momentum=\(k_{s}\)](p),
        };
    \end{feynman}
\end{tikzpicture}
}\\
    \subfloat[(b) ]{\begin{tikzpicture}
    \begin{feynman}
        \vertex (a){\(N\)};
        \vertex [right= of a](b);
        \vertex [right= of b](c);
        \vertex [right= of c](d);
        \vertex [right= of d](e){\(N\)};
        \vertex [below= of a](j){\(N\)};
        \vertex [right= of j](i);
        \vertex [right= of i](f);
        \vertex [right= of f](g);
        \vertex [right= of g](h){\(N\)};
        \vertex [below right= of g] (p){\(S\)};

        \diagram*{
        (a)--[fermion,momentum=\(p_{1}\)](b)--[fermion](c)--[fermion](d)[fermion]--[fermion,momentum=\(p_{3}\)](e),
        (j)--[fermion,momentum=\(p_{2}\)](i)--[fermion](f)--[fermion](g)--[fermion,momentum=\(p_{4}\)](h),
        (f)--[scalar,momentum'=\(k-k_{s}\),edge label=\(\pi\)](c),
        (g)--[scalar,momentum'=\(k_{s}\)](p),
        };
    \end{feynman}
\end{tikzpicture}
}\quad
    \subfloat[(d)]{\begin{tikzpicture}
    \begin{feynman}
        \vertex (a){\(N\)};
        \vertex [right= of a](b);
        \vertex [right= of b](c);
        \vertex [right= of c](d);
        \vertex [right= of d](e){\(N\)};
        \vertex [below= of a](j){\(N\)};
        \vertex [right= of j](i);
        \vertex [right= of i](f);
        \vertex [right= of f](g);
        \vertex [right= of g](h){\(N\)};
        \vertex [below right= of i] (p){\(S\)};

        \diagram*{
        (a)--[fermion,momentum=\(p_{1}\)](b)--[fermion](c)--[fermion](d)[fermion]--[fermion,momentum=\(p_{3}\)](e),
        (j)--[fermion,momentum=\(p_{2}\)](i)--[fermion](f)--[fermion](g)--[fermion,momentum=\(p_{4}\)](h),
        (f)--[scalar,momentum'=\(k-k_{s}\),edge label=\(\pi\)](c),
        (i)--[scalar,momentum'=\(k_{s}\)](p),
        };
    \end{feynman}
\end{tikzpicture}
}\quad
\subfloat[(e)]{\begin{tikzpicture}
    \begin{feynman}
        \vertex (a){\(N\)};
        \vertex [right= of a](b);
        \vertex [right= of b](c);
        \vertex [right= of c](d);
        \vertex [below = 0.9cm of c](q);
        \vertex [below = 0.9cm of d](r){\(S\)};
        \vertex [right= of d](e){\(N\)};
        \vertex [below= of a](j){\(N\)};
        \vertex [right= of j](i);
        \vertex [right= of i](f);
        \vertex [right= of f](g);
        \vertex [right= of g](h){\(N\)};
        \vertex [below right= of i] (p);

        \diagram*{
        (a)--[fermion,momentum=\(p_{1}\)](b)--[fermion](c)--[fermion](d)[fermion]--[fermion,momentum=\(p_{3}\)](e),
        (j)--[fermion,momentum'=\(p_{2}\)](i)--[fermion](f)--[fermion](g)--[fermion,momentum'=\(p_{4}\)](h),
        (f)--[scalar,edge label=\(\pi\)](c),
        (q)--[scalar,momentum=\(k_{s}\)](r),
        };
    \end{feynman}
\end{tikzpicture}}

    \caption{Feynman diagrams showing production of scalar fields via t-channel nucleon-nucleon bremsstrahlung. $N = n, p$.}
    \label{Feynman diagram}
\end{figure}

In the OPE approximation, the nucleon-nucleon bremsstrahlung process has both t and u channel contributions. Each channel has five subdiagrams (see Fig.~\ref{Feynman diagram}). Four of these diagrams have a scalar-nucleon-nucleon interaction and are labeled as ($a$), ($b$), ($c$), and ($d$) (($a'$), ($b'$), ($c')$, and ($d'$) for u-channel). The fifth diagram has a scalar-pion-pion interaction and is labeled as ($e$) (($e'$) for u-channel). The interaction of the CP-even scalar with the nucleon and pion is given by the effective Lagrangian~\cite{Chivukula:1989ds,Fradette:2017sdd},

\begin{equation}\label{lagra}
    \mathcal{L} = \sin{\theta}S\left[-y_{hNN}\bar{N}N+\frac{4}{9v}(\partial_{\mu}\pi^{+}\partial^{\mu}\pi^{-}+\frac{1}{2}\partial_{\mu}\pi^{0}\partial^{\mu}\pi^{0})-\frac{5m^{2}_{\pi}}{3v}(\pi^{+}\pi^{-}+\frac{1}{2}\pi^{0}\pi^{0})\right]
\end{equation}
where $y_{hNN} \approx 10^{-3}$ is the would-be effective coupling of a light SM Higgs to nucleons~\cite{Shifman:1978zn,Cheng:1988im}, and $v \simeq 246$ GeV is the electroweak vacuum expectation value. Note that the interaction between the scalar and the pions is different from Ref.~\cite{Dev:2020eam}, which is only valid when the pions are on-shell. Using the Lagrangian \ref{lagra}, we can calculate the amplitude for the nucleon-nucleon bremsstrahlung as described in detail in Appendix~\ref{appendix:amplitude}. 

The calculation simplifies significantly by making use of the fact that the nuclei are non-relativistic. As has been pointed out in previous work~\cite{Dev:2020eam}, there is a cancellation in the amplitudes of the ($a$), ($b$), ($c$), and ($d$) diagrams (similarly for the ($a'$), ($b'$), ($c'$), and ($d'$) diagrams) in the non-relativistic limit, i.e., when approximating the denominator of the internal fermion propagators by $(m_N \omega)^{-1}$, where $m_N$ is the nucleon mass and $\omega$ the scalar energy. This requires including higher order terms in the expansion of the fermion propagators. Thus far, expansions to next to leading order the scalar mass, $m_s$, have been explored~\cite{Dev:2020eam,Hardy:2024gwy}, but they neglected the expansions in momentum of the scalar ($\boldsymbol{k}_{s}$) and nucleon ($\boldsymbol{p}_i)$~(see also Ref.~\cite{Fiorillo:2025zzx} which kept higher order terms similar to our approach in the context of neutron star cooling). In order to explore the target parameter space, we choose to keep the leading order terms in momentum of the scalar and nucleon. We find that these terms have a significant contribution in the lower mass regime (scalars lighter than $ 100$ MeV) as they increase the luminosity by an order of magnitude.

In the isospin symmetric limit we are working with, there are further cancellations in the calculation requiring us to expand the propagators to higher orders in $v_i/c$, where $v_i$ are the nuclei velocities. This is expected since in the isospin limit the medium is expected to only radiate scalars through its quadrupole moment~\cite{Rrapaj:2015wgs,Fiorillo:2025zzx}. 

In addition, we not only need to expand the fermion propagator, but also take into account the difference in 4-momentum going through the pion propagator between the different diagrams. Both these effects are not captured by relying on the soft-radiation approximation. The detailed calculation of amplitudes using the improved expansion of the propagator is discussed in Appendix~\ref{appendix:amplitude}. Non-negligible corrections to our calculation from in-medium effect, two-pion exchange and other mesons, effective nucleon mass, and multiple-nucleon scattering effect as considered in Refs.~\cite{Carenza:2019pxu,Fiorillo:2025zzx} have been neglected for simplicity.
Recently, Ref.~\cite{Hardy:2024gwy} considered the production of CP-even scalars in proto-neutron stars via resonant mixing with the in-medium longitudinal photon for $m_s < 10$ MeV and soft production via nucleon-nucleon bremsstrahlung for higher masses. We have not included this channel, since with the improved bremsstrahlung calculation, this contribution is subdominant at all masses of interest.

\begin{figure}
\centering
\begin{subfigure}{.45\textwidth}
  \centering
  \includegraphics[width=0.8\linewidth]{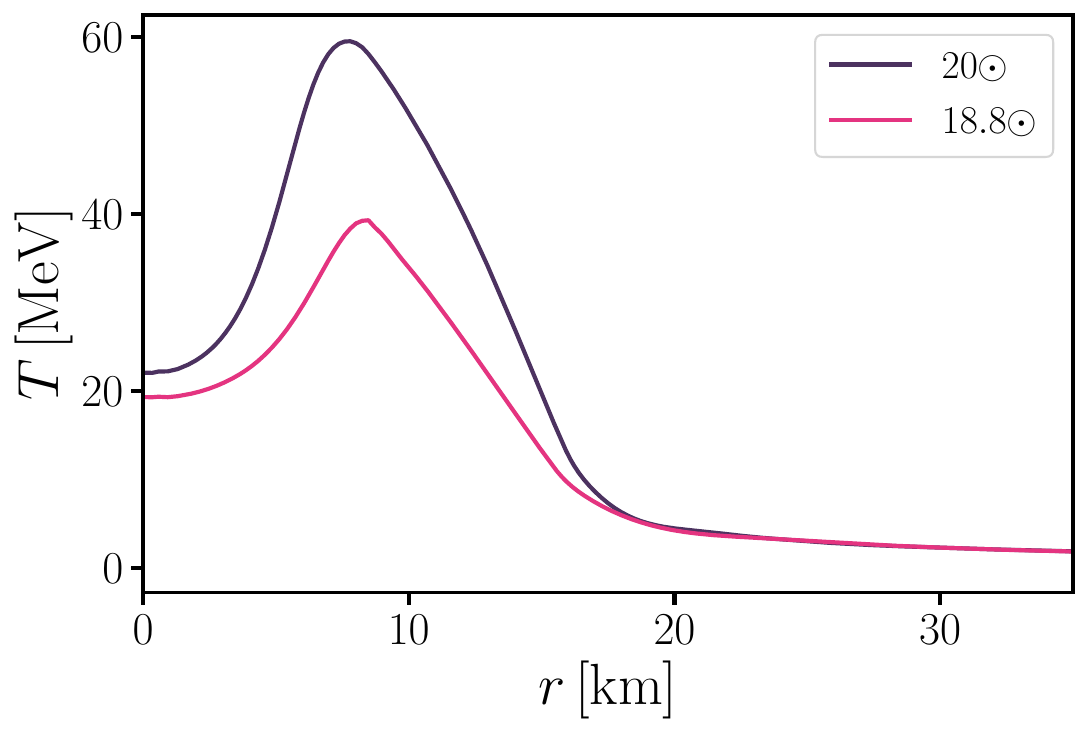}
\label{fig:temps}
\end{subfigure}
\begin{subfigure}{.45\textwidth}
  \centering
  \includegraphics[width=0.8\linewidth]{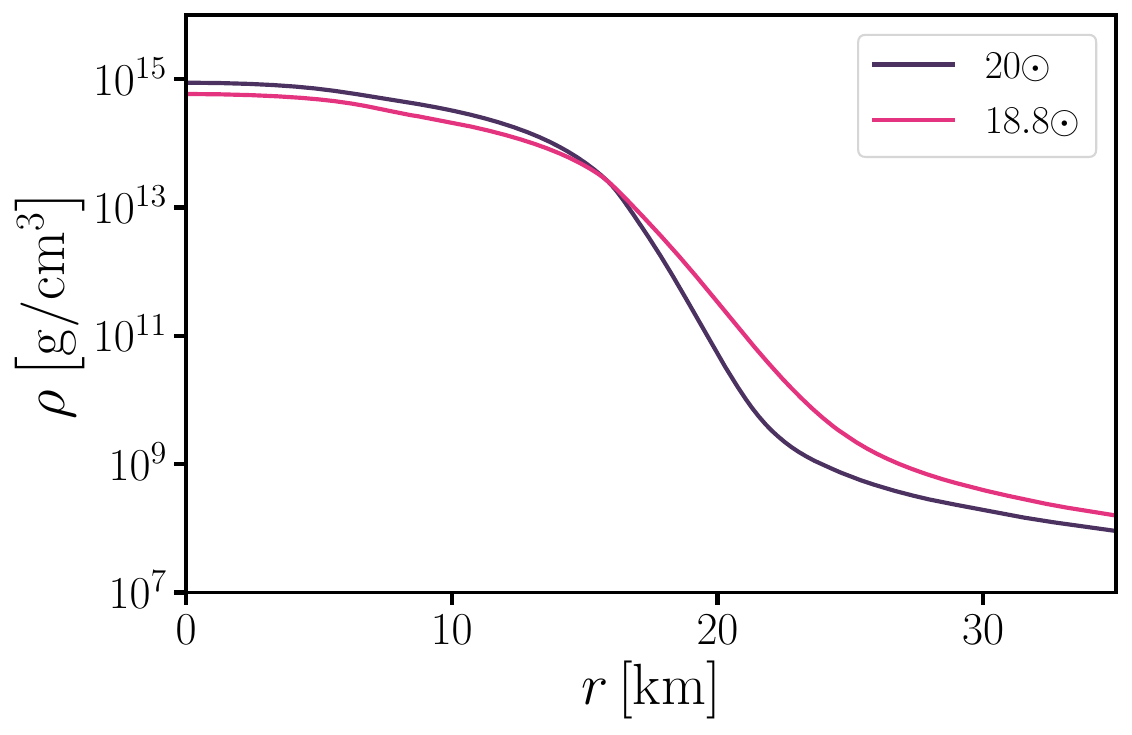}
    \label{fig:rhos}
\end{subfigure}
\caption{Density and Temperature profile of the SN at $t=1$s post bounce. The violet curve represents the hotter profile LS220-s20.0 with progenitor mass $20 \odot$ , and the pink curve represents the colder profile SFHo-18.8 with progenitor mass $18.8 \odot$.}
\label{fig:profiles}
\end{figure}
The instantaneous luminosity of the scalar per unit volume in the production frame is given by 
\begin{equation} \label{emiss}
\frac{dL}{dV}=\int d\Pi_N d\Pi_s  S\sum_{spins} |\mathcal{M}|^2 f_1 f_2 (2 \pi)^4\delta^4(p_1+p_2-p_3-p_4-k_s)\omega \, ,
\end{equation}
where $d\Pi_N$ and $d\Pi_s$ are the phase space factors for the four nucleons and the scalar, respectively. $\mathcal{M}$ is the amplitude for the $NN\rightarrow NNS$ process, which represents the neutron-neutron ($nn$), neutron-proton ($np$), and proton-proton ($pp$) processes. $S$ is a symmetry factor used to account for identical-particle initial and final states, where $S=1$ for $np$ processes and $S = 1/4$ for $nn$ and $pp$ processes. $f_1$ and $f_1$ are non-relativistic Maxwell-Boltzmann distributions for the incoming nucleons,
\begin{equation} \label{boltzdist}
f_i=\frac{n_B}{2}\left(\frac{2 \pi}{m_N T}\right)^{3/2}e^{-\mathbf{p_i}^2/2m_N T} \, ,
\end{equation}
where $n_{B}$ is the baryon number density and $T$ is the temperature of the location where the nucleons interact to produce the scalar.  Note that to simplify the calculation we have not included Fermi blocking terms in Eq.~\ref{emiss} as the matter in the SN core is only partially degenerate. Further simplification and calculation of the emissivity is performed similar to Refs.~\cite{Giannotti:2005tn,Dent:2012mx} and is discussed in Appendix~\ref{appendix:amplitude}.

We use the profile generated by the Garching 1D model SFHo-18.8 and LS220-s20.0 using the simulation by the PROMETHEUS VERTEX code~\cite{Garching}. Fig.~\ref{fig:profiles} shows these profiles. SFHo-18.8 is a colder profile with a maximum core temperature of $40 \,$MeV, and LS220-s20.0 is a hotter profile with a maximum core temperature of $60 \,$MeV. The progenitor masses are $18.8 \odot$ and $20 \odot$ for the models SFHo-18.8 and LS220-s20.0, respectively. These are towards the lower and upper limits of the allowed progenitor mass~\cite{Page:2020gsx}.

\section{Absorption and Decay}\label{sec:abs_and_decay}
\begin{figure}
    \centering
    \includegraphics[width=0.7\linewidth]{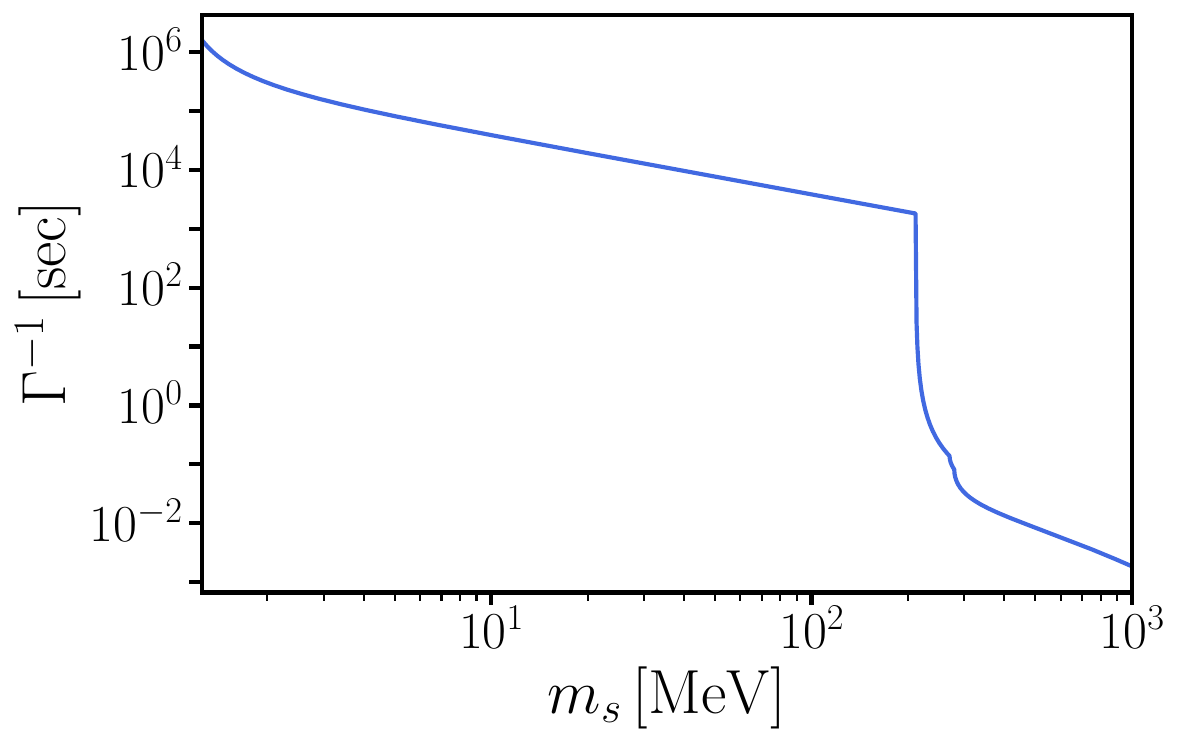}
    \caption{Lifetime of scalar $S$ as a function of their mass $m_s$ for mixing angle $\sin\theta = 10^{-7}$. }
    \label{fig:lifetime}
\end{figure}

After production in the SN core, the scalar can decay, or it can be reabsorbed via inverse nucleon-nucleon bremsstrahlung. If reabsorption or decay occurs near the core (which is the case for sufficiently large values of $\sin \theta$), the scalar's energy is merely redistributed locally, leading to trapping of the scalar and no reliable astrophysical bound. To properly apply various bounds in the following sections, we must therefore account for the effects of absorption and decay following production of the scalar in the core. First, we will discuss in general how to calculate the probability of re-absorption and decay for a scalar of a given mass and coupling produced at an initial radii $r$. However, as we will discuss, including the exact probability in our calculation would be numerically expensive, so, similarly to previous work, we will implement a simplified approach that is a good approximation to describe these effects. The relevant quantity we need to compute is the optical depth—the integrated probability that a scalar produced at radius $r_1$ is absorbed or decays before reaching radius $r_2$. The following decay channels are the most relevant to our parameter space,
\begin{align} \label{decays}
&\Gamma^0(S\rightarrow e^{-}e^{+})=\frac{m_s m_e^2\sin^2\theta}{8 \pi v^2}\left(1-\frac{4m_e^2}{m_s^2}\right)^{3/2}, \nonumber\\
&\Gamma^0(S\rightarrow \mu^{-}\mu^{+})=\frac{m_s m_\mu^2\sin^2\theta}{8 \pi v^2}\left(1-\frac{4m_\mu^2}{m_s^2}\right)^{3/2}, \nonumber\\
&\Gamma^0(S \rightarrow \pi^{+}\pi^{-})=\frac{\sin^2\theta}{324 \pi m_s v^2}\left(m_s^2+\frac{11}{2}m_{\pi^\pm}^2\right)^2\left(1-\frac{4m_{\pi^\pm}^2}{m_s^2}\right)^{1/2}, \nonumber\\
&\Gamma^0(S\rightarrow \pi^{0}\pi^{0})=\frac{\sin^2\theta}{648 \pi m_s v^2}\left(m_s^2+\frac{11}{2}m_{\pi^0}^2\right)^2\left(1-\frac{4m_{\pi^0}^2}{m_s^2}\right)^{1/2},
\end{align}
where the superscript $0$ denotes quantities evaluated in the rest frame of the scalar. We focus on scalars with mass $m_s > 2 m_e$, in which case they decay predominantly into $e^+e^-$ (with a small branching fraction into photons) until $m_s = 2 m_{\mu}$ at which point they decay predominantly into muons. This shift in decay from $e^+e^-$ to  $\mu^+\mu^-$ results in a sharp decrease in the lifetime of the scalar at $m_s = 2 m_{\mu}$ as seen in Fig.~\ref{fig:lifetime}. At $m_s=2m_{\pi}$, the pion decay channel opens, which further reduces the decay lifetime for $m_s > 2m_{\pi}$. In this work, we focus exclusively in the case where the primary decay channels are to the SM. Allowing for a large branching to an invisible dark sector species would lead to a stronger cooling bound but weaken the visible decay constraints we discuss later. A scalar mediator heavier than the dark matter mass is also highly constrained~\cite{Krnjaic:2015mbs}.

To properly characterize the decay probability, we account for both the scalar's boost and gravitational redshift. We define the  mean free path (MFP) of a particle, produced at radius $r$ with energy $\omega$, to be
\begin{equation} \label{declen}
\lambda_{dec}^{-1}(r,r',\omega)=\frac{\Gamma^{0}}{\gamma v}=\frac{m_s\Gamma^{0}}{|\mbf{k_s}|}=m_s\Gamma^{0} \left. \middle/ \sqrt{\frac{\eta(r)^2}{\eta(r')^2}\omega^2-m_s^2}\right. \, ,
\end{equation}
when it reaches radius $r'$, where $\eta$ is the gravitational lapse function \cite{Baumgarte_Shapiro_2010}, and $\mathbf{k_s}$ the momentum of the scalar at radius $r'$.

In addition to decaying, scalars can be reabsorbed via inverse bremsstrahlung, $NNS\rightarrow NN$. This process can be important at large couplings $\sin\theta$, as the nucleon density remains high up to approximately 30 km from the core center (see Fig.~\ref{fig:profiles}). The MFP of the scalar due to inverse bremsstrahlung is given by,
\begin{equation} \label{mfp}
\lambda^{-1}_{abs}(r',\omega)=\frac{1}{2 |\mbf{k_s}|}\int d\Pi_N f_1 f_2 S \sum_{spins} |\mathcal{M}(-k_s)|^2(2 \pi)^4 \delta^4(p_1+p_2-p_3-p_4+k_s) \, ,
\end{equation}
where the absorption amplitude, $\mathcal{M}(-k_s)$, can be obtained by reversing the scalar 4-momentum $k_s$ in the production amplitude. We can define an analogous mean free path to Eq.~\ref{declen} for the absorption, to take into account the impact of gravitational redshift.

The probability that a scalar produced at $r_1$ with energy $\omega$ reaches radius $r_2$ without being absorbed or decaying is encoded in the optical depth,
\begin{equation} \label{opacity}
\tau(r_1,r_2,\omega)=\int_{r_1}^{r_2}\lambda^{-1}( \omega)dr',
\end{equation}
with $\lambda^{-1} = \lambda_{dec}^{-1} + \lambda_{abs}^{-1}$. The equation above assumes that the scalar travels radially outward after being produced, but this is not necessarily the case.  Although it is possible to account for the angular dependence of the MFP~\cite{Balaji:2022noj}, it leads to a significant increase in computational time, especially for the case of re-absorption. For this reason, we will not consider the angular dependence explicitly and instead will treat the trapping due to reabsorption by using a black-body approximation as discussed in Section~\ref{sec:rafbound}. 

The impacts of the optical depth on the observed luminosity are most significant when the coupling is large, in which case the scalar can locally thermalize and has an approximately thermal spectrum with a temperature characteristic of the production region. We use this to simplify the optical depth calculation by using the energy-averaged MFP, given by, 
\begin{equation} \label{enavg}
\langle\lambda^{-1}\rangle(r,r')=\int_{m_s/\eta}^{\infty} \lambda^{-1}(r,r',\omega) \frac{\omega^2 \sqrt{\omega^2-m_s^2}\text{ }d\omega}{e^{\omega/T}-1} \left. \middle/\int_{m_s/\eta}^{\infty} \frac{\omega^2 \sqrt{\omega^2-m_s^2}\text{ }d\omega}{e^{\omega/T}-1}\right. ,
\end{equation}
where $\eta$ and $T$ are evaluated at the production location. Note that Eq.~\ref{mfp} for the absorption MFP depends only on the radius $r'$ where absorption occurs and the energy of the scalar at that radius, but the energy-averaging procedure introduces a dependence on the production radius $r$ through the assumed thermal spectrum at the production location. We use this energy averaged MFP to calculate the typical optical depths,
\begin{equation} \label{opacityavg}
\tau(r_1,r_2)=\int_{r_1}^{r_2}\langle \lambda^{-1}\rangle (r_1,r')dr',
\end{equation}
which depends only on the initial and final radii, instead of keeping track of the optical depth as a function of energy. This expression is used in all subsequent sections when dealing with processes that occur in or near the core.

When exploring signals from visible decays, we will be interested in scalars that decay far away from the core. In this case, a different simplification is possible. The expressions for decay in Eqs.~\ref{decays} and~\ref{declen} are much simpler than those for absorption in Eq.~\ref{mfp}, depending on radius only through the lapse functions. At these large radii, the lapse function approaches unity ($\eta(r') \approx 1$) so the optical depth integral can be approximated analytically and we can retain the explicit energy dependence in the MFP instead of relying on thermal averaging:

\begin{equation}
\tau_{dec}(r,R,\omega)=m_s \Gamma^0\int_{r}^{R} \frac{dr'}{\sqrt{\frac{\eta(r)^2}{\eta(r')^2}\omega^2-m_s^2}}\approx \frac{m_s \Gamma^0(R-r)}{\sqrt{\eta(r)^2\omega^2-m_s^2}}.
\end{equation}
It will be convenient to define
\begin{equation}\label{opacityapprox}
\widetilde{\tau}_{dec}(r,R,\omega)\equiv\frac{m_s\Gamma^0 (R-r)}{\sqrt{\eta(r)^2\omega^2-m_s^2}},
\end{equation}
for use in later sections where scalars decay well beyond the core region.

\section{SN1987a Cooling Bound}\label{sec:rafbound}

The core-collapse SN SN1987a in the Large Magellanic Cloud was the closest SN event that has occurred in the last century, and the first for which we were able to detect a neutrino signal. The Kamiokande, IMB and Baksan experiments observed approximately 24 neutrinos signals over a 10 second interval coincident with SN1987a~\cite{Kamiokande-II:1987idp,Bionta:1987qt,Alekseev:1987ej}. These observations are in agreement with the delayed explosion theory of core-collapse SN~\cite{Bethe:1985sox}. According to the theory, the energy deposited by the neutrinos revives the stalled shock waves after the core bounce, forming a new proto-neutron star (PNS). The newly formed PNS cools down by emitting neutrinos with energies of order 10 MeV. Simulations by Lattimer and Burrows provide strong evidence that the neutrino events observed for SN1987a  are due to the cooling of the PNS~\cite{Burrows:1986me,Burrows:1987zz,Burrows:1988ba}.

Any weakly interacting BSM particle produced inside the PNS can transport energy out of the core and cool the PNS. If the energy transported by the BSM particle is comparable or larger than that of the neutrinos, the neutrino emission duration and it's spectra would be substantially impacted, spoiling the agreement with the observed signal. The total energy transported by the neutrinos accounts for roughly $ \approx 2-3 \times 10^{53}~ \text{erg}$ of the PNS's gravitational binding energy. Based on this, the upper bound on the maximum luminosity that any new particle can have without affecting  the neutrino observations of 1987a is 
\begin{equation}
    L_{NP} \leq 3 \times 10^{52} ~\text{erg}/\text{s},
\end{equation}
which is popularly known as the Raffelt criterion or SN cooling bound~\cite{Raffelt:1996wa}. Using the improved production rate calculation, we use this upper limit to place bounds on the mixing angle $\sin{\theta}$ and mass $m_s$ of the CP-even scalar. This SN cooling bound has  been previously explored in the context of light vector bosons~\cite{Dent:2012mx,Kazanas:2014mca,Rrapaj:2015wgs,Chang:2016ntp}, axions~\cite{Raffelt:1987yt,Turner:1987by,Burrows:1988ah,Brinkmann:1988vi,Ishizuka:1989ts,Giannotti:2005tn,Carenza:2019pxu,Choi:2021ign}, ALP's~\cite{Jaeckel:2017tud,Lee:2018lcj}, and scalars~\cite{Krnjaic:2015mbs,Dev:2020eam,Balaji:2022noj,Hardy:2024gwy}.

For sufficiently low values of $\sin\theta$, scalars produced in the core are highly unlikely to undergo any further nuclear interactions once emitted, allowing them to free-stream. We calculate the minimum $\sin\theta$ for which this free-streaming luminosity would violate the Raffelt criterion. Integrating Eq.~\ref{emiss} over the core volume and equating it to $3 \times 10^{52}~ \text{erg}/\text{s}$ gives the lower bound on $\sin\theta$,
\begin{equation}\label{Llower}
    L_{lower} = 4\pi \int_{0}^{R_c}\eta(r)^{2}\frac{dL}{dV}r^2dr =3 \times 10^{52} \text{erg}/s,
\end{equation}
where $R_c$ is the core radius, which we take to be 30 km. By this radius the nuclear density has dropped significantly, allowing us to safely neglect scalar production beyond this point. Note that we have included the lapse functions ($\eta$) to correct for the gravitational redshift since the Raffelt criterion is based on the observed instantaneous luminosity. 

For significantly large values of $\sin\theta$, interactions between scalar and SM particles become strong enough to trap the scalar in the core region via absorption and decay. This trapping determines the maximum value for $\sin\theta$ value that can be constrained using the Raffelt criterion. Modifying Eq~\ref{emiss}, and taking the trapping and gravitational redshift into account, the observed instantaneous luminosity per unit volume would be given by,
\begin{equation}\label{emisstrap}
    \frac{dL_{trap}}{d\Omega dV} = \eta(r)^2\frac{e^{-\tau}}{4\pi}\frac{dL}{dV},
\end{equation}
where we have re-introduced the angular dependence of the emission because the optical depth $\tau$ depends on the propagation direction of the scalar. The optical depth measures the probability that a scalar produced at a certain radii $r$ would propagate to a cutoff radius $R_f$ without being re-absorbed or decaying so that the energy carried by the scalar is not redistributed back to the core or neutrinos. Following Ref.~\cite{Chang:2016ntp}, we take this radius to be 100 km. Given the computational cost of the multi-dimensional integration already required for the production (and absorption) cross-section, we adopt a simplified approach described below to handling re-absorption, in which the full angular dependence is not explicitly tracked~(see Ref.~\cite{Balaji:2022noj} for an analysis that kept the full angular dependence and Ref.~\cite{Hardy:2024gwy} for an approximation that is somewhat similar to the one we employ).

\begin{figure}
\centering
\includegraphics[width=0.7\linewidth]{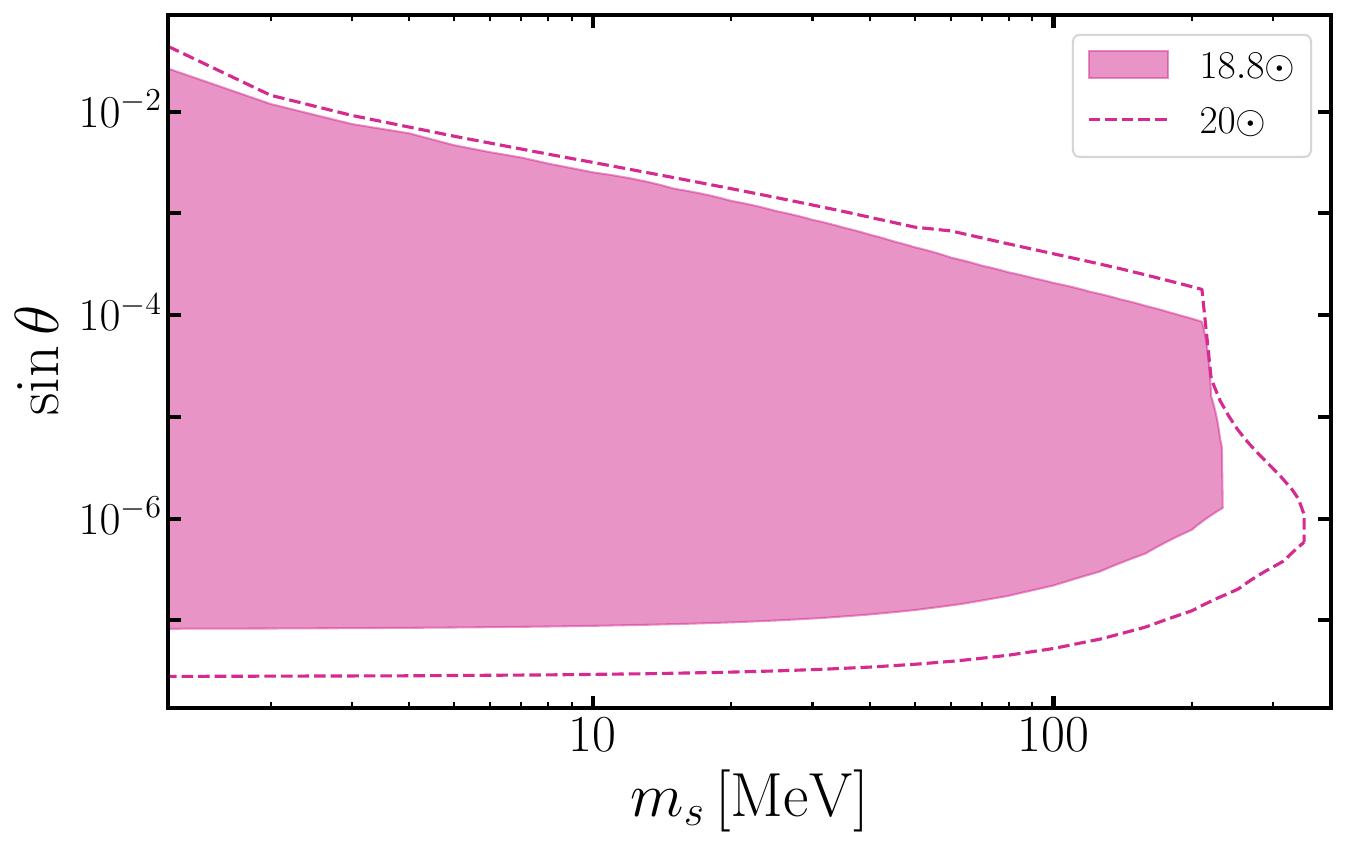}
\caption{SN cooling bound on scalar $S$. The shaded region represents the bounds using the colder profile, whereas the region inside the dotted line represents the bounds using the hotter profile.}
\label{fig:raff-cooling}
\end{figure}

The MFP depends quadratically on the number density of baryons, which drops extremely sharply as the radius increases (see Fig.~\ref{fig:profiles}).  This means that the optical depth changes very abruptly once one is outside the densest part of the core. We can thus define a decoupling radius, where the optical depth (considering only absorption) for a radial trajectory is equal to $2/3$~\cite{Raffelt:2001kv}. Effectively all scalars produced inside of this decoupling radius will be reabsorbed, while almost all scalars produced at or above the decoupling radius will free-stream, as long as their trajectory doesn't go back inside the decoupling radius. These conditions allow us to treat the decoupling radius as a black-body.  The luminosity of bosons produced by the black-body (at the decoupling radius), taking into account the correction from gravitational redshift, is given by,
\begin{equation} \label{bblum}
L_{b}=\eta(R_{b})^2\frac{\pi^3 }{30 }g^{*}(q) R_{b}^2 T_{b}^4 ,
\end{equation}
where
\begin{equation} \label{gstar}
g^{*}(q)=g_i\frac{15}{\pi^4}\int_{q/\eta}^\infty \frac{x(x^2-q^2) dx}{\exp(x)-1},
\end{equation}
is the mass dependent correction incurred when the bosons are not fully relativistic ($g_i=1$ is the number of internal degrees of freedom of our scalar). $R_{b}$ is the black-body radius, and $T_{b}$ is the temperature at the black-body radius. As described above, the black-body radius was chosen as the location where $\tau_{abs}(R_b,R_f) = 2/3$, that is,
\begin{equation}
    \tau_{abs}(R_{b},R_f) = \int_{R_{b}}^{R_f}\langle\lambda_{abs}^{-1}\rangle(R_{b},r') dr' = \frac{2}{3}.
\end{equation}
The blackbody emission accounts for the scalars produced at radii $\lesssim R_b$. Due to the large nuclear density even above $R_b$, there is non-negligible production of the scalar beyond the black-body radius. Taking this into consideration, we add the volume emission term to the blackbody luminosity and define the instantaneous luminosity at the trapping regime as,
\begin{equation}\label{Ltrap}
   L_{trap}=L_{b}e^{-\tau_{dec}(R_{b},R_{f})}+2 \pi\int_{R_{b}}^{R_c}r^{2}dr\int^{\infty}_{m_{s}/\eta} d\omega ~\eta(r)^{2}e^{-\tilde{\tau}_{dec}(r,R_f,\omega)}\frac{dL}{dVd\omega}. 
\end{equation}
The factor $e^{-\tau_{dec}(R_b,R_f)}$ accounts for the decay of the scalars produced at the black body radius $R_{b}$, and we have only considered emissions over a solid angle of $2\pi$ to avoid trajectories that would go back into the black-body. Note that, because the particles are produced at radii $r \ll R_f$, and the decay probability per unit length is approximately constant, it is a good approximation to use the optical depth due to decays assuming a radial trajectory. We also evaluated the trapping luminosity using the prescription in Ref.~\cite{Hardy:2024gwy}, in which the average optical depth was calculated assuming that half of the particles travel radially outward and half radially inward. This prescription gave results similar to ours.

In Fig.~\ref{fig:raff-cooling}, we present our cooling bounds. The shaded region represents the bounds using the colder profile, whereas the region inside the dotted line represents the bounds using the hotter profile. As expected, the bound from the hotter profile is more constraining than that from the colder profile, due to the larger production rate resulting from the higher temperature.

\section{Low-Energy Supernovae}\label{sec:low_energy}

Energy escaping from the core may decay within the progenitor star, contributing to the SN's observed energy. Core-collapse SN explosions span a wide range of energies. While typical SN energy is on the order of $10^{51}$ erg, observations and reconstructions of certain Type II-P SN indicate that the explosion energy of these core-collapse SN can range down to $10^{50}$ erg~\cite{Murphy:2019eyu}. The light curve of a Type II-P SN has a plateau shape, hence the P in Type II-P. The duration and brightness of the plateau phase are sensitive to the progenitor radius, the ejected mass, the explosion energy, and the amount of ejected ${}^{56}$Ni mass~\cite{Pejcha:2015pca,Muller:2017bdf,Kozyreva:2018uxe,Goldberg:2019ktf,Murphy:2019eyu}. The decay of  ${}^{56}$Co (daughter nucleus of ${}^{56}$Ni) to ${}^{56}$Fe heats the expanding debris of the SN, extending the plateau of the light curve. LE-SN with a ${}^{56}$Ni mass of roughly $10^{-3}M_\odot$,  which is more than 10 times lower than the typical core-collapse SN, have explosion energies on the order of $10^{50}$ erg~\cite{Valenti:2009ps,Spiro_2014}.

Such LE-SN provide a valuable probe of weakly coupled hidden sector production, and the parameter spaces of such scenarios have been studied in the context of ALP's~\cite{Caputo:2022mah} and sterile neutrinos~\cite{Chauhan:2023sci}.  Energy injected into the mantle by exotic species is constrained to be less than that inferred from Type II-P light-curves, which we set at $10^{50}$ergs.  We use the Garching group's SFH0-18.8 model to model the LE-SN progenitor core.  This model is the coldest of the Garching groups muonic models, with properties in agreement with theoretical models of LE-SN~\cite{Caputo:2022mah}. We set the radius of our progenitor at $R_{env}=3\times10^7$ km, and take the mantle region to be between $R_f$ and $R_{env}$. $R_f$ is the cutoff radius defined in Section~\ref{sec:rafbound}.

\begin{figure}
\centering
\includegraphics[width=0.7\textwidth]{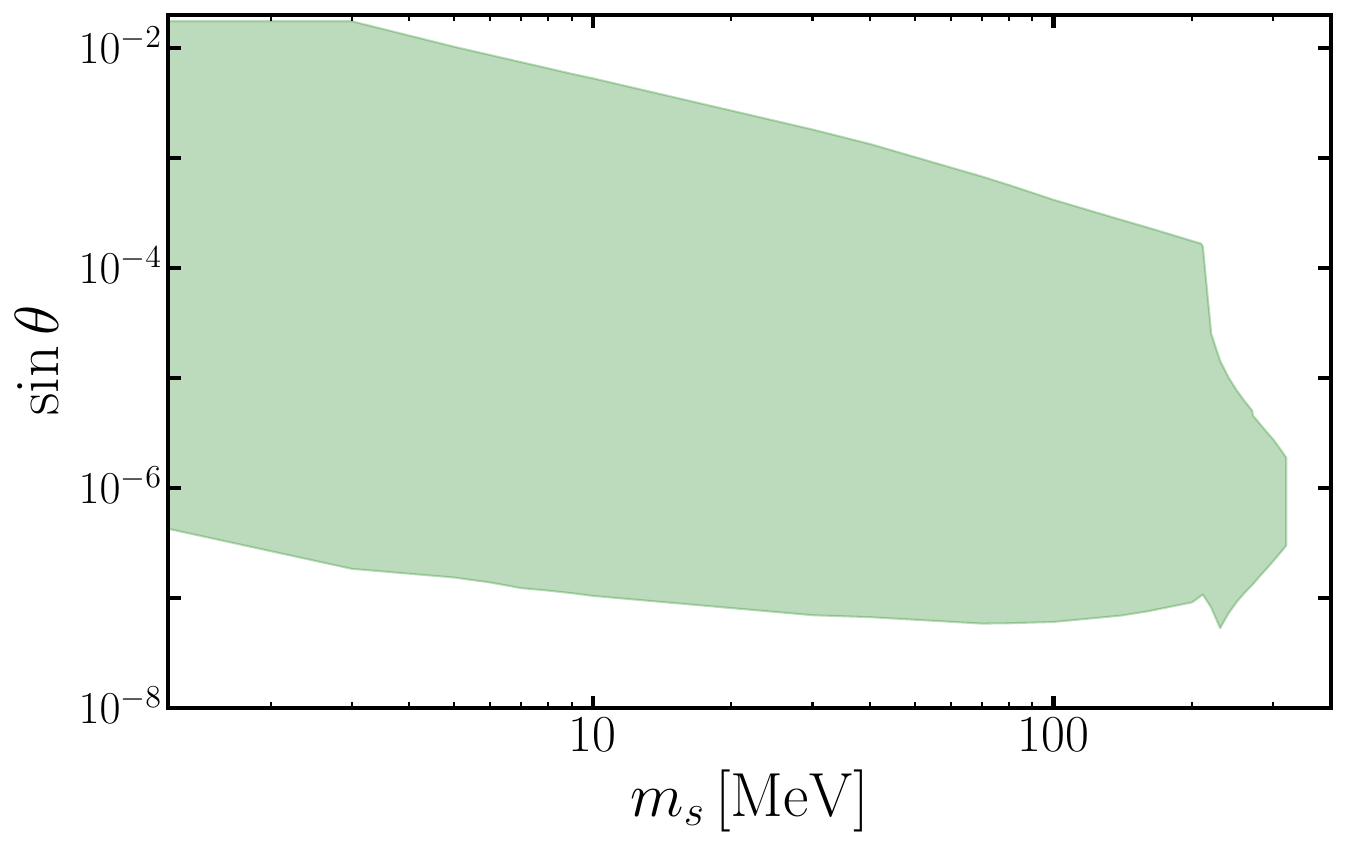}
\caption{Low-energy supernova explosion bound on scalar $S$.}
\label{fig:lesncold}
\end{figure}

The energy deposited in the mantle region is given by,
\begin{equation}\label{le-sn}
    E_{dep}=f_{dep}\int dt~ 4\pi\int^{R_c}_{0}~r^{2}dr\int^{\infty}_{m_s/\eta}~ d\omega~ \frac{dL}{dVd\omega}\left(e^{-\tilde{\tau}_{dec}(r,R_f,\omega)}-e^{-\tilde{\tau}_{dec}(r,R_{env},\omega)}\right).
\end{equation}
Even if the scalar decays in the mantle, the energy of the decay products may not be deposited in the mantle. Muons and pions produced in the scalar decay can further decay to produce neutrinos which can travel outside the mantle without depositing their energy. To account for the fraction of the decay energy deposited into the mantle, we define $f_{dep}$ as the fraction of the energy of the scalars that goes into stable visible particles (not including neutrinos which free-stream carrying the energy away). We find that $f_{dep} = 7/20$ for $2m_{\mu}<m_s<2m_{\pi}$. For $m_s > 2m_{\pi}$, the scalars decay primarily into charged and neutral pions. Since charged pions decay into muon plus neutrino, the latter further decays into neutrinos plus positrons, and so only a small fraction of the original charged pion energy goes into visible particles.  For simplicity, we conservatively assume that only the scalar decays to neutral pions deposit energy into the mantle, hence $f_{dep}=1/3$ for $m_s > 2m_{\pi}$. For large values of $\sin \theta$, the scalars can become trapped in the core, limiting energy deposition in the mantle and setting an upper bound on $\sin \theta$. Modeling the trapping regime as in the previous section using the blackbody approximation, the energy deposited in the mantle is given by, 
\begin{align}\label{le-sntrap}
    E_{dep,trap}&= f_{dep}\int dt~L_{b}\left(e^{-\tau_{dec}(R_b,R_f)}-e^{-\tau_{dec}(R_b,R_{env})}\right) \nonumber \\
   &+ f_{dep} \int dt~ 2\pi\int^{R_c}_{0}~r^{2}dr\int^{\infty}_{m_s/\eta}~ d\omega~ \frac{dL}{dVd\omega}\left(e^{-\tilde{\tau}_{dec}(r,R_f,\omega)}-e^{-\tilde{\tau}_{dec}(r,R_{env},\omega)}\right).
\end{align}
Equating Eq.~\ref{le-sn} and Eq.~\ref{le-sntrap} to $10^{50}$ erg gives the LE-SN bound shown in Fig.~\ref{fig:lesncold}.

\section{Positron Bound}\label{sec:positron_bound}

Various balloon-based missions in the 1970s have detected 511 keV gamma ray flux from the galactic center region corresponding to the electron-positron annihilation~\cite{johnson1972spectrum,haymes1975detection,leventhal1978detection}. Recently, the SPI gamma ray spectrometer on the INTEGRAL satellite has improved upon these measurements significantly~\cite{Siegert:2015knp}. Based on its measurements, the estimated rate of galactic positron annihilation should be no larger than $\sim 4\times10^{43} ~\text{s}^{-1}$~\cite{Prantzos:2010wi}. If we assume that the positron production rate and annihilation rate are in equilibrium, this can be taken as the galactic positron production rate. Any astrophysical source that produces positrons at a rate greater than $4\times10^{43}~\text{s}^{-1}$ would be in tension with the observation by INTEGRAL. Scalars produced in the PNS would predominantly decay into electron-positron pairs if in the mass range 1 to 211 MeV, thus contributing to the galactic positron population. As a result, a conservative bound can be placed on the scalar particles based on the observation of the 511 keV galactic gamma ray flux.

In order to contribute to the galactic positron population, the scalar has to escape the PNS outer layer and gaseous envelope. The composition and size of the outer layer depend on the type of SN event. Since we have considered only Type II SN for other bounds in this work, we consider only Type II SN to contribute towards the galactic positron population. Including Type I SN would yield a slightly stronger bound, as they happen less frequently than Type II SN and typically have smaller escape radii ($R_{esc}$). However, this parameter space is already constrained by the 1987a cooling and LE-SN bound. Following the calculation in Ref.~\cite{DeRocco:2019njg}, we take the $R_{esc}$ to be $10^9$ km. Type II SN occur in our galaxy with an average rate of 2 SN/century~\cite{Adams:2013ana}. Based on this, in order to not exceed the galactic positron production rate, the maximum number of allowed positrons per SN event is $\sim 6.3\times 10^{52}$. Thus, the regions of parameter space in which scalars produced in a Type II SN would produce over $10^{53}$ positrons are excluded. Note that considering the morphology of the observed 511 keV gamma ray flux should lead to a slightly stronger limit~\cite{Calore:2021lih}.

\begin{figure}
\centering
\includegraphics[width=0.7\linewidth]{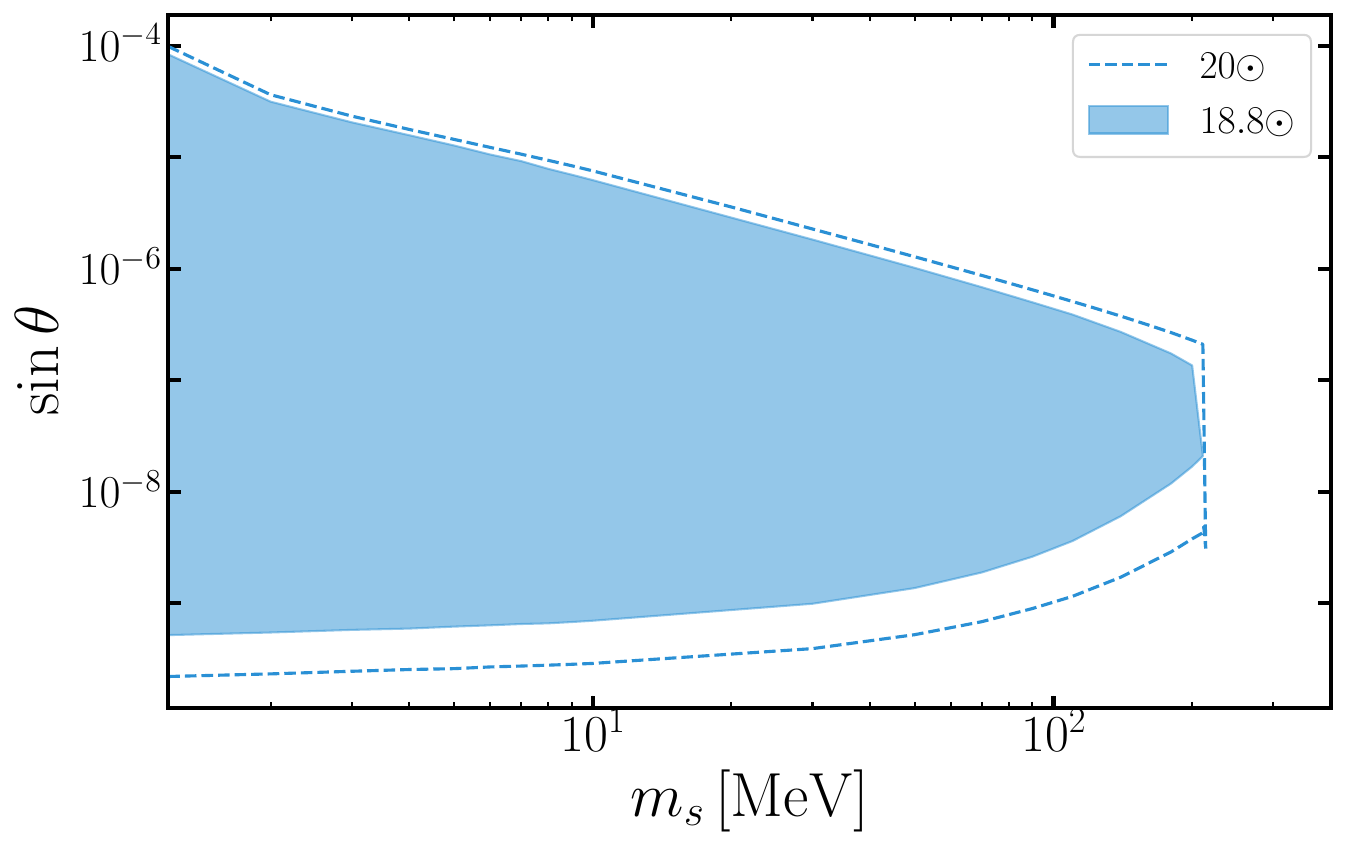}
\caption{Positron bound on scalar $S$. The shaded region represents the bounds  using the colder profile, whereas the region inside the dotted line represents the bounds using the hotter profile.}
\label{fig:posbound}
\end{figure}

The number of scalars produced per unit volume and time is given by,
\begin{equation} \label{posprod}
\frac{dN}{dtdV}=\int d\Pi_N d\Pi_s  S \sum_{spins}|\mathcal{M}|^2 f_1 f_2 (2 \pi)^4\delta^4(p_1+p_2-p_3-p_4-k_s).
\end{equation}
Integrating this over the volume of the core and the time scale of scalar production ($\sim 10$ s) gives the total number of scalars produced, which is equivalent to the the number of positrons yielded by their decays. If all of these scalars travel beyond $R_{esc}$, they will contribute to the observed positron flux. Equating this to $10^{53}$ gives a lower bound on $\sin \theta$,

\begin{equation}\label{positron lower}
    N_{e^{+}lower}=\int~dt\int 4\pi r^{2} \eta(r)\frac{dN}{dtdV}dr= 10^{53}.
\end{equation}
Higher values of $\sin\theta$ will result in the production of more positrons, violating the galactic positron production rate. However, for a significantly higher value of $\sin \theta$, the decay length of the scalar becomes small, such that it decays into positrons before reaching $R_{esc}$. This results in an upper bound on the $\sin\theta$. Since this occurs in a regime with larger $\sin \theta$, we also allow for the possibility that reabsorption could be relevant and determine the black-body radius for that $\sin \theta$ (which is set to 0 if $\tau_{abs}$ is always less than $2/3$), and model the luminosity similarly to how we treated the trapped regime for the cooling bound. The number of bosons produced by the black-body per unit time, taking into account the correction from gravitational redshift, is given by, 
\begin{equation}
    \frac{dN_{b}}{dt} = \eta(R_{b})\frac{\pi^3 }{30 }g_{N}^{*}(q) R_{b}^2 T_{b}^3,
\end{equation}
where
\begin{equation} 
g_{N}^{*}(q)=g_i\frac{15}{\pi^4}\int_{q/\eta}^\infty \frac{(x^2-q^2) dx}{\exp(x)-1}.
\end{equation}
Finally, the total number of positrons produced in the trapping regime is given by,
\begin{equation}\label{positron trap}
    N_{e^{+}trap}=\int \frac{dN_{b}}{dt}e^{-\tau(R_b,R_{esc})}dt ~+\int dt~2 \pi\int_{R_{b}}^{R_c}r^2dr\int^{\infty}_{m_{s}/\eta} d\omega~\eta(r)e^{-\tau_{dec}(r,R_{esc},\omega)}\frac{dN}{dtdVd\omega}.
\end{equation}
Using Eq.~\ref{positron lower} and Eq.~\ref{positron trap}, we find the positron bound, which is presented in Fig.~\ref{fig:posbound}. There is a sharp cutoff in the bound at $m_s = 2m_{\mu}$, when the scalar predominantly decays into muons, resulting in a sharp decline in the lifetime of the scalar (see Fig.~\ref{fig:lifetime}), pushing the decays to be before $R_{esc}$.

Positrons escaping from the neutron star mantle may interact with electrons to produce photons via pair production and bremsstrahlung. If these interactions are efficient enough, hot dense plasma will form outside the mantle leading to the formation of a fireball. The formation and subsequent expansion of the plasma can significantly deplete the number of positrons that would eventually contribue towards the  511 keV line observed by SPI~\cite{DeRocco:2019njg}. Thus, in order to assess the validity of the positron upper bound derived here, it is necessary to determine if a fireball will form in any part of the positron bound. Following the analysis of \cite{Diamond:2021ekg,Diamond:2023scc}, we evaluate the conditions for fireball formation by calculating the optical depth of the decay products surrounding the PNS. We find that a fireball is formed only in a small part of the positron bound. In that region, the majority of the energy would be converted to photons and thus also constrained by the non-observation of gamma-rays~\cite{Diamond:2023scc}. We also find that these regions are already constrained by cooling or the LE-SN bound.

\section{Hadrophilic scalar}\label{sec:nucleophilic}
\begin{figure}[t!]
\centering
\includegraphics[width=0.7\textwidth]{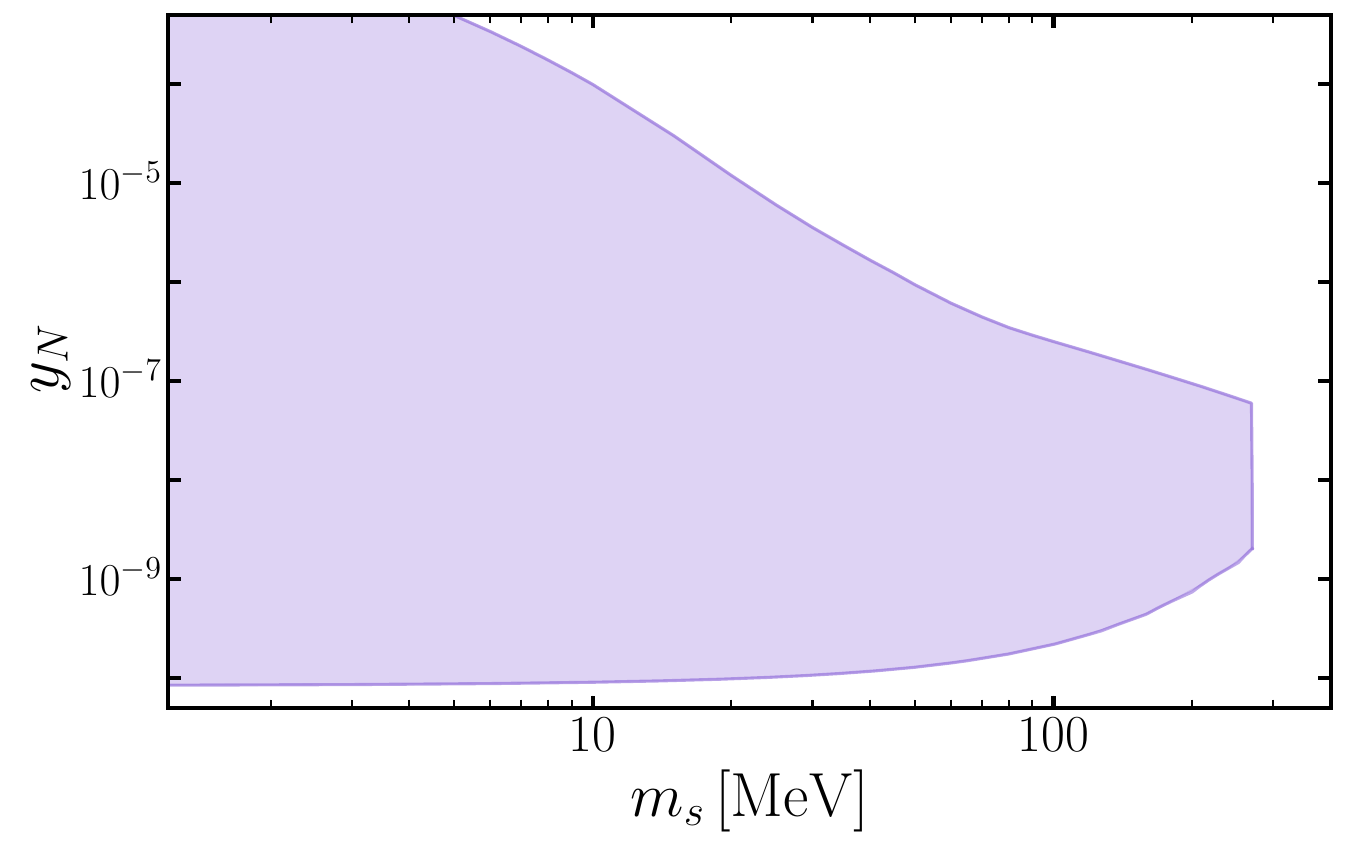}
\caption{SN cooling bound on hadrophilic scalar $S_N$ using the colder profile.}
\label{fig:np-cold}
\end{figure}

We will extend the supernova bounds to a hadrophilic scalar $S_{N}$ that couples to nucleons via the interaction,
\begin{equation}\label{npL}
    \mathcal{L} = -y_{N}S_{N}\bar{N}N,
\end{equation}
without generating a tree-level coupling to leptons. The hadrophilic interaction in Eq.~\ref{npL} can be generated by a scalar-gluon interaction,
\begin{equation}\label{scalar-gluon}
    \mathcal{L} = \frac{\epsilon}{v}\frac{\alpha_{s}}{12\pi}S_{N}G^{a}_{\mu\nu}G^{a\mu\nu},
\end{equation}
where the scalar-gluon coupling $\epsilon$ and nucleophilic coupling $y_{N}$ are related by~\cite{Gunion:1989we},
\begin{equation}
    y_{N} = \frac{2}{3}\frac{\epsilon~m_{N}}{b~v}.
\end{equation}
Here $b$ is the first coefficient of the QCD $\beta$-function.

The scalar-gluon interaction generates a scalar-pion interaction~\cite{Chivukula:1989ds},
\begin{equation}\label{npLpion}
   \mathcal{L}= S_{N}\frac{\epsilon}{bv}\left[\frac{4}{3}(\partial_{\mu}\pi^{+}\partial^{\mu}\pi^{-}+\frac{1}{2}\partial_{\mu}\pi^{0}\partial^{\mu}\pi^{0})-2m^{2}_{\pi}(\pi^{+}\pi^{-}+\frac{1}{2}\pi^{0}\pi^{0})\right].
\end{equation}
The interaction of hadrophilic scalars with pions differs from the Higgs-mixed-scalar $S$ case in Eq.~\ref{lagra} because the scalar-gluon interaction solely generates the interactions and does not have light quark contributions.  Using the Lagrangians~\ref{npL} and~\ref{npLpion}, we can calculate the production rate of the hadrophilic scalar following the same approach we developed for the Higgs mixed scalar (see Appendix~\ref{appendix:amplitude}) by replacing the nucleon interaction strength $\sin\theta ~y_{hNN}$ by $y_{N}$, and the pion interaction strength $\sin\theta\mathcal{A}_{\pi}$ by $\tilde{\mathcal{A}}_{\pi}$. See Appendix~\ref{appendix:amplitude} for the definitions of $\mathcal{A}_{\pi}$ and $\tilde{\mathcal{A}}_{\pi}$.

Unlike the Higgs-mixed-scalar scenario, there are no leptonic decays in the case of hadrophilic scalars. The important processes for the trapping regime are reabsorption and decays to photons and pions. The decay rate to pions, calculated using Lagrangian~\ref{npLpion}, is given by,
\begin{equation}
    \Gamma^{0}(S_{N}\rightarrow \pi^{+}\pi^{-}) = \frac{y_{N}^{2}}{16\pi}\frac{(m^{2}_{s}+m^{2}_{\pi})^{2}}{m_{s}m_{N}^{2}}\left(1-\frac{4m^{2}_{\pi}}{m_{s}^{2}}\right)^{1/2}.
\end{equation}

The scalar-gluon interaction in Eq.~\ref{scalar-gluon} generates a scalar-photon interaction. Following the calculations in Ref.~\cite{Leutwyler:1989tn}~(see also Ref.~\cite{Delaunay:2025lhl} for a more general hadrophilic scalar model), the scalar-photon interaction is given by, 
\begin{equation}
    \mathcal{L}=\frac{y_{N}}{m_{N}}\frac{\alpha}{8\pi}\frac{14}{3}S_{N}F_{\mu\nu}F^{\mu\nu},
\end{equation}
where we have considered virtual loops of pions and kaons and have assumed the number of heavy fermions to be 3, i.e, $b=9$. Using the above interaction, the decay rate to a photon is given by,
\begin{equation}
    \Gamma^{0}(S_{N}\rightarrow\gamma\gamma)= \frac{y^{2}_{N}}{m^{2}_{N}}\left(\frac{\alpha}{8\pi}\right)^{2}\left(\frac{14}{3}\right)^{2}\frac{m^{3}_{s}}{4\pi}.
\end{equation}

Using Eq.~\ref{Llower} and Eq.~\ref{Ltrap} for the luminosity, we find the 1987a cooling bound on the hadrophilic scalar, shown in Fig.~\ref{fig:np-cold}. There is a sharp feature in the bound when $m_s=2m_\pi$ when the decay channel into pions opens leading to much more efficient trapping at large couplings.

\section{Conclusion and Outlook}\label{sec:conclusion}

\begin{figure}
\centering
\includegraphics[width=1.0\linewidth]{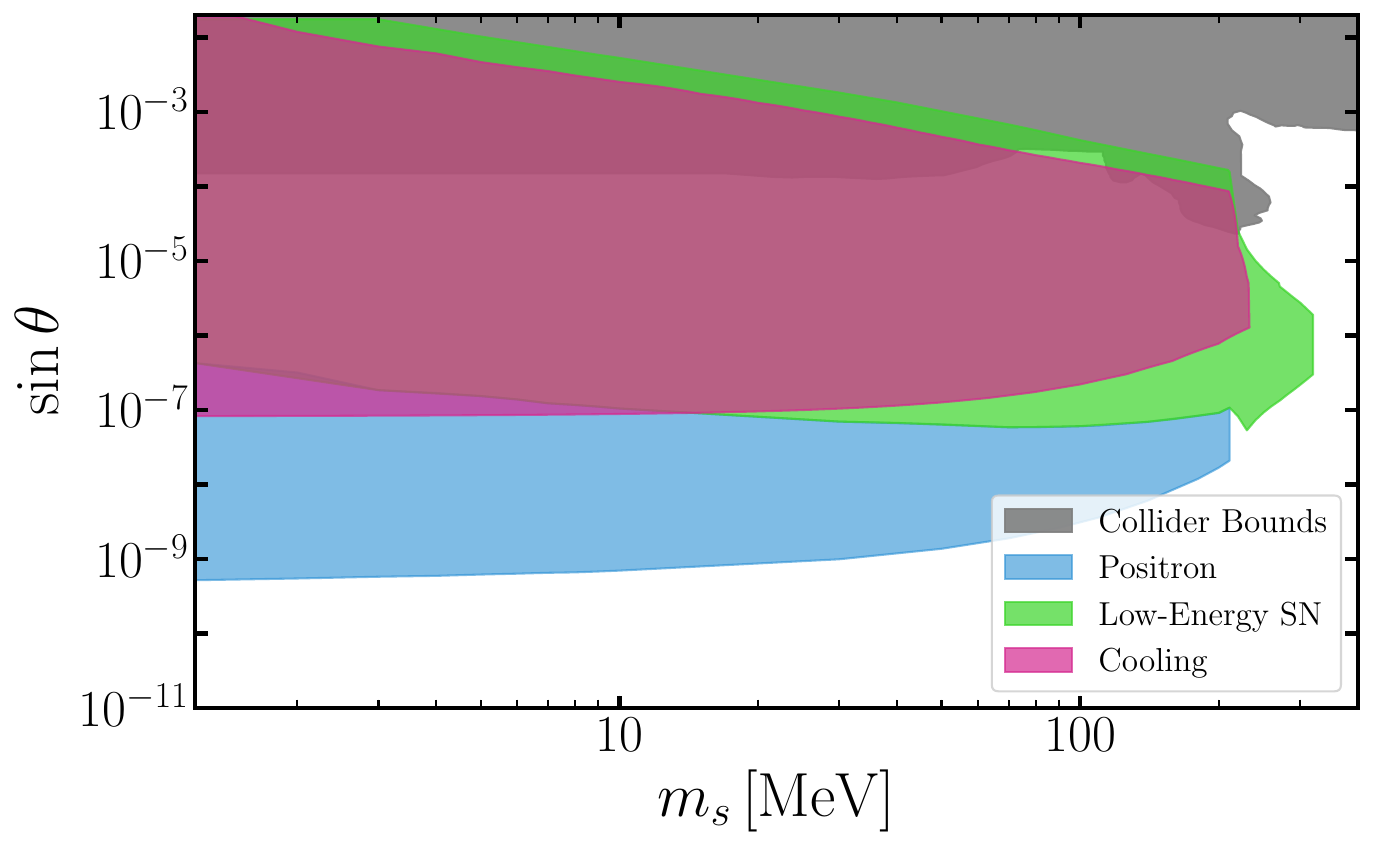}
\caption{SN bounds on scalar mass $m_s$ and mixing angle $\sin\theta$ obtained for the colder SN profile. The shaded pink region represents the SN cooling bound, whereas the blue shaded region represents the positron bound. The shaded green region represents the LE-SN bound. The grey shaded region represents the existing collider bounds from Ref.~\cite{Batell:2022dpx}}
\label{fig:full-cold}
\end{figure}

In this work, we revisited the supernova bounds on a CP-even scalar that mixes with the Higgs and considered for the first time the signatures due to its decays back to standard model particles. We summarize our bounds in Fig.~\ref{fig:full-cold}, which were derived using the colder profile considered in our study, making those more conservative. The cooling bound, complemented by the positron bound and LE-SN bound, helps us probe the $\sin\theta$ down to $\sim 10^{-9}$, which is more than five orders of magnitude below the existing collider bounds~\cite{Batell:2022dpx} (gray shaded region) for a MeV scale CP-even scalar as seen in Fig.~\ref{fig:full-cold}.
The SN cooling bound (shaded pink region in Fig.~\ref{fig:full-cold}) has been improved by more than an order of magnitude relative to previous bounds due to improvements made in the production rate calculation in the (OPE) approximation and the treatment of the trapping regime. In the non-relativistic limit, at leading order, the amplitudes of the scalar production via the OPE approximation cancel out. We improve the production rate by retaining next-to-leading-order terms proportional to the nucleon and scalar momenta. We also include the correct interaction between the scalar and the pion, which differs from Ref.~\cite{Dev:2020eam}. This changes the contributions to production and absorption from the internal pion propagator.

We place the positron bound by restricting the number of positrons produced from the decay of the scalar to be less than $10^{53}$ so that it doesn't violate the 511 keV galactic gamma ray flux observed by the INTEGRAL satellite. The blue shaded region in Fig.~\ref{fig:full-cold} represents the positron bound. As shown in the figure, we can constrain the $\sin\theta$ value to two orders of magnitude below the cooling bound using the positron bound. This represents the region in the parameter space where not enough scalars are produced to violate the luminosity limit (cooling bound), but result in excessive positron production via decay.

The shaded green region in Fig.~\ref{fig:full-cold} represents the LE-SN bound. This corresponds to the region in the parameter space where the total energy deposited by the scalar into the mantle exceeds the energy of the lowest energetic core-collapse SN observed, taken to be $10^{50}$ erg. Since this is a weaker constraint compared to the cooling bound, we can probe slightly above the upper limit from the cooling bound using the LE-SN bound.

In addition, we also computed the cooling bounds for a simple hadrophilic scalar model. In the free-streaming regime, the production is almost identical to the higgs portal case, except for a negligible difference coming from the pion vertex. However, in the trapping regime, we find a stronger bound for the hadrophilic case, since for the same coupling to nuclei, the decay width of the hadrophilic scalar is much smaller due to the absence of decays to leptons.

The combination of astrophysical and collider bounds covers over nine orders of 
magnitude in coupling, probing the majority of the parameter space motivated by 
dark matter models, and demonstrating the strong complementarity between supernova 
signals and collider searches for MeV-scale particles. Given the inherent 
uncertainties in astrophysical constraints --- illustrated by the difference between 
bounds derived using the colder and hotter profiles --- each bound should be 
understood as carrying an order-of-magnitude level uncertainty. The OPE 
approximation introduces a similar theoretical uncertainty, and extending the 
calculation beyond OPE would be a valuable check on the robustness of our results. 
Nonetheless, since the different bounds overlap across a wide range of couplings, 
carry independent systematics, and were derived using the more conservative cold 
profile, we expect the bounds presented here to be robust at the level of a factor 
of $2$--$3$ in the coupling.

\section{Acknowledgments} 

We would like to thank Hans-Thomas Janka for providing the numerical profiles used to simulate the progenitor core. The work of GMT and MJ is partially supported in part by the  National Science Foundation under Grant Number PHY-2412828. The works of SL was supported by the U.S. Department of Energy under Award No. DESC0009959.  The support and resources from the Center for High Performance Computing at the University of Utah are gratefully acknowledged. 

\appendix

\section{Amplitude Calculation}
\label{appendix:amplitude}

First, we show the calculation for the proton-proton bremsstrahlung $p + p \rightarrow p + p + S$ process. The amplitudes of t-channel diagrams (see Fig. \ref{Feynman diagram}) for this process are,
\begin{align}
    \mathcal{M}_a &=-g_{pp\pi}^{2}\frac{\sin \theta y_{hNN}}{k^2-m_{\pi}^2}\frac{\bar{u}(p_4)\gamma^5u(p_2)\bar{u}(p_3)(\slashed{p_3}+\slashed{k_s}+m_N)\gamma^5u(p_1)}{(p_3+k_s)^2-m_N^2}, \nonumber \\
    \mathcal{M}_b &=-g_{pp\pi}^{2}\frac{\sin \theta y_{hNN}}{(k-k_s)^2-m_{\pi}^2}\frac{\bar{u}(p_4)(\slashed{p_4}+\slashed{k_s}+m_N)\gamma^5u(p_2)\bar{u}(p_3)\gamma^5u(p_1)}{(p_4+k_s)^2-m_N^2}, \nonumber \\
    \mathcal{M}_c &=-g_{pp\pi}^{2}\frac{\sin \theta y_{hNN}}{k^2-m_{\pi}^2}\frac{\bar{u}(p_4)\gamma^5u(p_2)\bar{u}(p_3)\gamma^5(\slashed{p_1}-\slashed{k_s}+m_N)u(p_1)}{(p_1-k_s)^2-m_N^2}, \nonumber \\
    \mathcal{M}_d &= -g_{pp\pi}^{2}\frac{\sin \theta y_{hNN}}{(k-k_s)^2-m_{\pi}^2}\frac{\bar{u}(p_4)\gamma^5(\slashed{p_2}-\slashed{k_s}+m_N)u(p_2)\bar{u}(p_3)\gamma^5u(p_1)}{(p_2-k_s)^2-m_N^2}, \nonumber\\
    \mathcal{M}_e & = g_{pp\pi}^{2} \frac{\sin \theta \mathcal{A}_{\pi}(k,k_{s})}{k^2-m_{\pi}^2}\frac{\bar{u}(p_4)\gamma^5u(p_2)\bar{u}(p_3)\gamma^5u(p_1)}{(k-k_{s})^{2}-m^{2}_{\pi}}.
\end{align}
Here $p_{i}$($i=1,2,3,4$) represents the momentua of the nucleons, and $k_{s}$ represents the momentum of scalar S. $k = p_{2}-p_{4}$ is the momentum exchanged in the t-channel. The  $p-p-\pi$ interaction is given by  $g_{pp\pi}\pi^{0}\bar{p}i\gamma_{5}p$, where $g_{pp\pi} =(2m_{N}/m_{\pi})f_{pp}$ is the effective coupling, with $f_{pp} \approx 1$. $\mathcal{A}_{\pi}(k,k_{s})$ is the interaction strength of the scalar with the pion given by,
\begin{equation}
    \mathcal{A}_{\pi}(k,k_s) = \frac{1}{v}\left[\frac{4}{9}k(k-k_{s})-\frac{5}{3}m_{\pi}^{2}\right].
\end{equation}
Note that this interaction strength between the scalar and the pion differs from Ref.~\cite{Dev:2020eam}. The u-channel diagrams take similar forms and can be obtained by transforming $p_{3}\leftrightarrow p_{4}$ everywhere (which necessitates the transformation $k\rightarrow l = p_{2}-p_{3}$) and multiplying the diagrams by an overall minus sign to account for fermion spin statistics. The amplitudes for the hadrophilic scalar can be obtained by replacing the nucleon interaction strength $\sin\theta~ y_{hNN}$ by $y_N$ and the pion interaction strength $\sin \theta~ \mathcal{A}_{\pi}$ by,
\begin{equation}
    \tilde{\mathcal{A}}_{\pi} = \frac{y_{N}}{m_{N}}\left[2k(k-k_{s})-3m^{2}_{\pi}\right].
\end{equation}

In order to simplify the expression, we further expand the propagator term as follows,
\begin{align}\label{prop appro 1}
    \frac{1}{(p_i\pm k_s)-m_N^2} &= \frac{1}{\pm 2 p_i\cdot k_s+m_s^2} = \frac{1}{\pm2(E_N \omega-\mathbf{p_i}\cdot \mathbf{k_s})+m_s^2} \nonumber \\
    & = \frac{1}{\pm2m_{N}(1+\mathbf{p^{2}_{i}}/2m_N^2)\omega\mp2\mathbf{p_{i}}\cdot \mathbf{k_{s}}+m^{2}_{s}} \nonumber \\
    & \approx \pm \frac{1}{2m_{N}\omega}\left(1-\frac{\mathbf{p^{2}_{i}}}{2m_{N}^{2}} + \frac{\mathbf{p_{i}}\cdot\mathbf{k_{s}}}{m_{N}\omega} + \frac{(\mathbf{p_{i}}\cdot\mathbf{k_{s}})^{2}}{m^{2}_{N}\omega^{2}} \mp \frac{m^{2}_{s}} {2m_{N}\omega}\right).
\end{align}
Here, $+$ corresponds to the expansion for diagrams (a) and (c), and $-$ corresponds to the expansion for diagrams (b) and (d). In the above equation, we have used the non-relativistic expansion for the nucleon energy $E_{N} \approx m_{N}(1+\mathbf{p_{i}^{2}}/2m^{2}_{N})$.
If we keep only the leading term, i.e.,
\begin{equation}
    \frac{1}{(p_i\pm k_s)-m_N^2} \approx \frac{1}{\pm2m_{N}\omega},
\end{equation}
the sum of the four amplitudes with emission from external legs $\mathcal{M}_{a,b,c,d}$ (also $ \mathcal{M}_{a',b',c',d'} $) cancel out, requiring the inclusion of higher order terms. In the current literature, the expansion has only been made to $\mathcal{O}(m^2_{s}/m_{N}\omega)$, the last term in Eq.~\ref{prop appro 1}.
However, in the low-mass regime,  contributions from terms dependent on the scalar momentum will have larger impacts than terms dependent on the scalar mass.  We include the relevant $\mbf{k_s}$ dependent expansion terms in Eq. \ref{prop appro 1}.  Note that we have kept terms which are of $\mathcal{O}(\mathbf{p}_{i}^{2}/m^{2}_{N})$, due to further cancellation which occurs when we sum the amplitudes $\mathcal{M}_{a,b,c,d}$ and use energy-momentum conservation. The scalar four-momentum also appears in the pion propagator of diagrams $(b)$ and $(d)$.  We  Taylor expand these propagators as follows,
\begin{align}\label{prop appro 2}
    \frac{1}{(k-ks)^2-m_{\pi}^2} &=\frac{1}{k^2-m_\pi^2-2 k\cdot k_s+m_{s}^2} \nonumber \\
    &\approx\frac{1}{k^2-m_\pi^2}\left(1-\frac{m_s^2-2k \cdot k_s}{k^2-m_{\pi}^2}\right).
\end{align}

 Since there are no cancellations in the (e) diagram, we choose to neglect the scalar momentum that appears in its denominator. To evaluate the spinor contribution to the amplitude square, we use the following kinematic relations~\cite{Dent:2012mx}, 
\begin{align}
    p_1\cdot p_3 &= m_N^2-\frac{1}{2}m_s^2-\frac{1}{2}k^2+k\cdot k_s \approx m_N^2-\frac{1}{2}k^2, \nonumber \\
    p_1\cdot p_4 &= m_N^2-\frac{1}{2}m_s^2-\frac{1}{2}l^2+l\cdot k_s \approx  m_N^2-\frac{1}{2}l^2,\nonumber \\
    p_2\cdot p_3 &=m_N^2-\frac{1}{2}l^2, \nonumber \\
    p_2\cdot p_4 &=m_N^2-\frac{1}{2}k^2 ,\nonumber \\
    p_3\cdot p_4 &=m_N^2-\frac{1}{2}k^2-\frac{1}{2}l^2+k \cdot l .
\end{align}
We will now make the approximations $k^{2} \approx -|\mathbf{k}^{2}|, l^{2} \approx -|\mathbf{l}^{2}|, (k\cdot l) \approx - (\mathbf{k} \cdot \mathbf{l})$ and $\mathbf{k}^{2}, \mathbf{l}^{2}, (\mathbf{k} \cdot \mathbf{l}) \gg m^{2}_{s}, \omega^{2} $ in the remaining calculations. Using the approximations for the propagators in Eq.~\ref{prop appro 1} and Eq.~\ref{prop appro 2} we find,
\begin{align}\label{eq:mt}
    |\mathcal{M}_{t}|_{pp}^2  =~&|\mathcal{M}_{a+b+c+d}|_{pp}^{2}
     = g_{pp\pi}^{4}\frac{(\sin{\theta}y_{hNN})^2}{4m^{2}_{N}\omega^{2}} \left[\frac{1}{2m_{N}^{2}}(\boldsymbol{p}_{3}^{2}+\boldsymbol{p}_{4}^{2}-\boldsymbol{p}_{2}^{2}-\boldsymbol{p}_{1}^{2}) +\frac{2m_{s}^{2}}{m_{N}\omega} \right. \nonumber\\
    &\left. +\frac{(\boldsymbol{p}_{1}+\boldsymbol{p}_{2}-\boldsymbol{p}_{3}-\boldsymbol{p}_{4})\cdot\boldsymbol{k}_{s}}{m_{N}\omega}+ \frac{(\boldsymbol{p}_{1}\cdot\boldsymbol{k}_{s})^{2}+(\boldsymbol{p}_{2}\cdot\boldsymbol{k}_{s})^{2}-(\boldsymbol{p}_{3}\cdot\boldsymbol{k}_{s})^{2}-(\boldsymbol{p}_{4}\cdot\boldsymbol{k}_{s})^{2}}{m_{N}^{2}\omega^{2}}\right. \nonumber \\
    & \left. -\frac{(2 k\cdot k_s-m_s^2)(\mbf{k}\cdot \mbf{k_s}+m_s^2)}{m_{N}\omega(\mathbf{k}^{2}+m_{\pi}^{2})}\right]^{2}\frac{16 m^{2}_{N}\mathbf{k}^{4}}{(\mathbf{k}^{2}+m^{2}_{\pi})^{2}}.
\end{align}
The factor of $16m^{2}_{N}\mathbf{k}^{4}$ represents the contribution arising from the spinors. Using conservation of momentum and energy,
\begin{equation}
    \boldsymbol{p}_{1}+\boldsymbol{p}_{2} = \boldsymbol{p}_{3} + \boldsymbol{p}_{4}+ \boldsymbol{k}_{s},
\end{equation}
\begin{equation}
     \frac{\boldsymbol{p}_{1}^{2}}{2m_{N}}  + \frac{\boldsymbol{p}_{2}^{2}}{2m_{N}}  =   \frac{\boldsymbol{p}_{3}^{2}}{2m_{N}} + \frac{\boldsymbol{p}_{4}^{2}}{2m_{N}} +\omega,
\end{equation}
Eq.~\ref{eq:mt} becomes,
\begin{align}
    |\mathcal{M}_{t}|_{pp}^2 = ~& g_{pp\pi}^{4} \frac{(\sin{\theta}y_{hNN})^2}{4m^{2}_{N}\omega^{2}} \left[\frac{2m^{2}_{s}}{m_{N}\omega}-\frac{m^{2}_{s}}{m_{N}\omega}-\frac{(2 k\cdot k_s-m_s^2)(\mbf{k}\cdot \mbf{k_s}+m_s^2)}{m_{N}\omega(\mathbf{k}^{2}+m_{\pi}^{2})}\right.  \nonumber \\
    & \left. + \frac{(\boldsymbol{p}_{1}\cdot\boldsymbol{k}_{s})^{2}+(\boldsymbol{p}_{2}\cdot\boldsymbol{k}_{s})^{2}-(\boldsymbol{p}_{3}\cdot\boldsymbol{k}_{s})^{2}-(\boldsymbol{p}_{4}\cdot\boldsymbol{k}_{s})^{2}}{m_{N}^{2}\omega^{2}} \right]^{2}
    \frac{16 m^{2}_{N}\mathbf{k}^{4}}{(\mathbf{k}^{2}+m^{2}_{\pi})^{2}}.
\end{align}
Using energy and momentum conservation leads to a large cancellation, resulting in the $-m^{2}_S/m_{N}\omega$ term, justifying the retention of the $\mathcal{O}(\mathbf{p}_{i}^{2}/m^{2}_{N})$ terms in the propagator expansion earlier. The total amplitude for $p+p \rightarrow p+p+S$ process can be written as,
\begin{align}
    |\mathcal{M}|_{pp}^2 &= |\mathcal{M}_{t} + \mathcal{M}_{u} + \mathcal{M}_{e} + \mathcal{M}_{e'}|_{pp}^{2} \nonumber \\
   & = |\mathcal{M}_{t+u}|_{pp}^2 + |\mathcal{M}_{e+e'}|_{pp}^{2} + 2(|\mathcal{M}_{t}\mathcal{M}_{e}| +|\mathcal{M}_{u}\mathcal{M}_{e}|+|\mathcal{M}_{t}\mathcal{M}_{e'}|+|\mathcal{M}_{u}\mathcal{M}_{e'}|)_{pp} \nonumber \\
   & = |\mathcal{M}_{t+u}|_{pp}^2 + |\mathcal{M}_{e+e'}|_{pp}^{2} + 2|\mathcal{M}^{2}_{tuee'}|_{pp}.
\end{align}.

All terms in the amplitude square can be calculated similarly to $|\mathcal{M}_{t}|_{pp}^2$, which we summarize here,

\begin{align}
 |\mathcal{M}_{t+u}|_{pp}^2 &=~ g_{pp\pi}^{4}\frac{\sin^{2}{\theta}16 m^{2}_{N}}{4m^{2}_{N}} \left[\frac{\mathbf{k}^{4}}{(\mathbf{k}^{2}+m^{2}_{\pi})^{2}}\mathcal{T}^{2}_{1} + \frac{\mathbf{l}^{4}}{(\mathbf{l}^{2}+m^{2}_{\pi})^{2}}\mathcal{T}^{2}_{2}
    +\frac{\mathbf{k}^{2}\mathbf{l}^{2}-2(\mathbf{k}\cdot \mathbf{l})^{2}}{(\mathbf{k}^{2}+m^{2}_{\pi})(\mathbf{l}^{2}+m^{2}_{\pi})} \mathcal{T}_{1}\mathcal{T}_{2}\right],\nonumber \\
    \nonumber\\
 |\mathcal{M}_{e+e'}|_{pp}^{2} &=~4g_{pp\pi}^{4}\sin^{2}{\theta} \left[\frac{\mathbf{k}^{4}}{(\mathbf{k}^{2}+m^{2}_{\pi})^{4}}\mathcal{A}^{2}_{\pi}(\mathbf{k}^{2}) + \frac{\mathbf{l}^{4}}{(\mathbf{l}^{2}+m^{2}_{\pi})^{4}}\mathcal{A}^{2}_{\pi}(\mathbf{l}^{2})\right. \nonumber\\
 &\left.+\frac{(\mathbf{k}^{2}\mathbf{l}^{2}-2(\mathbf{k}\cdot \mathbf{l})^{2})}{(\mathbf{k}^{2}+m^{2}_{\pi})^{2}(\mathbf{l}^{2}+m^{2}_{\pi})^{2}} \mathcal{A}_{\pi}(\mathbf{k}^{2})\mathcal{A}_{\pi}(\mathbf{l}^{2})\right], \nonumber \\
 \nonumber \\
 2|\mathcal{M}^{2}_{tuee'}|_{pp} &=-g_{pp\pi}^{4}\frac{\sin^{2}{\theta}16m_{N}}{2m_{N}} \left[ \frac{\mathbf{k}^{4}}{(\mathbf{k}^{2}+m^{2}_{\pi})^{3}}\mathcal{T}_{1} \mathcal{A}_{\pi}(\mathbf{k}^{2}) + \frac{\mathbf{l}^{4}}{(\mathbf{l}^{2}+m^{2}_{\pi})^{3}}\mathcal{T}_{2}\mathcal{A}_{\pi}(\mathbf{l}^{2})\right. \nonumber\\
&\left.+\frac{1}{2}\frac{\mathbf{k}^{2}\mathbf{l}^{2}-2(\mathbf{k}\cdot \mathbf{l})^{2}}{(\mathbf{k}^{2}+m^{2}_{\pi})(\mathbf{l}^{2}+m^{2}_{\pi})^{2}}\mathcal{T}_{1}\mathcal{A}_{\pi}(\mathbf{l}^{2}) +\frac{1}{2}\frac{\mathbf{k}^{2}\mathbf{l}^{2}-2(\mathbf{k}\cdot \mathbf{l})^{2}}{(\mathbf{k}^{2}+m^{2}_{\pi})^{2}(\mathbf{l}^{2}+m^{2}_{\pi})}\mathcal{T}_{2}\mathcal{A}_{\pi}(\mathbf{k}^{2})\right], 
\end{align}
where $\mathcal{T}_{1}$, $\mathcal{T}_{2}$, and $\mathcal{A}_{\pi}$ are,
\begin{align}\label{simplifying terms}
    &\mathcal{T}_{1} =-\frac{y_{hNN}}{\omega}\left[\frac{m^{2}_{s}}{m_{N}\omega}-\frac{(2 k\cdot k_s-m_s^2)(\mbf{k}\cdot \mbf{k_s}+m_s^2)}{m_{N}\omega(\mathbf{k}^{2}+m_{\pi}^{2})} + \frac{\mathcal{R}}{m^{2}_{N}\omega^{2}}\right], \nonumber\\
    &\mathcal{T}_{2} = -\frac{y_{hNN}}{\omega}\left[\frac{m^{2}_{s}}{m_{N}\omega}-\frac{(2 l\cdot k_s-m_s^2)(\mbf{l}\cdot \mbf{k_s}+m_s^2)}{m_{N}\omega(\mathbf{l}^{2}+m_{\pi}^{2})}+ \frac{\mathcal{R}}{m^{2}_{N}\omega^{2}}\right], \nonumber \\
    &\mathcal{R} = (\boldsymbol{p}_{1}\cdot\boldsymbol{k}_{s})^{2}+(\boldsymbol{p}_{2}\cdot\boldsymbol{k}_{s})^{2}-(\boldsymbol{p}_{3}\cdot\boldsymbol{k}_{s})^{2}-(\boldsymbol{p}_{4}\cdot\boldsymbol{k}_{s})^{2}, \nonumber \\
     &\mathcal{A}_{\pi}(\mathbf{k}^{2}) = -\frac{1}{v}\left[\frac{4}{9}\mathbf{k}^{2}+\frac{5}{3}m_{\pi}^{2}\right], \nonumber \\
      &\mathcal{A}_{\pi}(\mathbf{l}^{2}) = -\frac{1}{v}\left[\frac{4}{9}\mathbf{l}^{2}+\frac{5}{3}m_{\pi}^{2}\right].
\end{align}
Adding all the terms, the amplitude square can be written most compactly as follows,
\begin{align} \label{mpp}
\sum_{spins}|\mathcal{M}_{pp}|^2 =~&
4 g_{pp\pi}^{4} \sin^2 \theta\left[\frac{\mbf{k}^4}{(\mbf{k}^2+m_{\pi}^2)^2}\left(\mathcal{T}_1-\frac{\mathcal{A}_{\pi}(\mathbf{k}^{2})}{\mbf{k}^2+m_{\pi}^2}\right)^2\right.
\left.+\frac{\mbf{l}^4}{(\mbf{l}^2+m_{\pi}^2)^2}\left(\mathcal{T}_2-\frac{\mathcal{A}_{\pi}(\mathbf{l^{2}})}{\mbf{l}^2+m_{\pi}^2}\right)^2\right.\nonumber\\
&\left.+\frac{\mbf{k}^2\mbf{l}^2-2(\mbf{k}\cdot \mbf{l})^2}{(\mbf{k}^2+m_{\pi}^2)(\mbf{l}^2+m_{\pi}^2)}\left(\mathcal{T}_1-\frac{\mathcal{A}_{\pi}(\mathbf{k^{2}})}{\mbf{k}^2+m_{\pi}^2}\right)\left(\mathcal{T}_2-\frac{\mathcal{A}_{\pi}(\mathbf{l}^{2})}{\mbf{l}^2+m_{\pi}^2}\right)\right].
\end{align}
To calculate $|\mathcal{M}|_{np}^{2}$ and $|\mathcal{M}|_{nn}^{2}$ we make use of the relationship $g_{nn\pi} = - g_{pp\pi} $ and $g_{np\pi} = - \sqrt{2}g_{pp\pi} $ given by isospin invariance. Since $g_{pp\pi} = - g_{nn\pi} $, $\sum_{spins}|\mathcal{M}|^{2}_{pp} = \sum_{spins}|\mathcal{M}|^{2}_{nn} $. Using  $g_{np\pi} = - \sqrt{2}g_{pp\pi}$ we get,
\begin{flalign} \label{mnp}
\sum_{spins}|\mathcal{M}_{np}|^2 =&~
4g_{pp\pi}^{4} \sin^2 \theta\left[\frac{\mbf{k}^4}{(\mbf{k}^2+m_{\pi}^2)^2}\left(\mathcal{T}_1-\frac{\mathcal{A}_{\pi}(\mathbf{k}^{2})}{\mbf{k}^2+m_{\pi}^2}\right)^2\right.
\left.+\frac{4\mbf{l}^4}{(\mbf{l}^2+m_{\pi}^2)^2}\left(\mathcal{T}_2-\frac{\mathcal{A}_{\pi}(\mathbf{l}^{2})}{\mbf{l}^2+m_{\pi}^2}\right)^2\right.\nonumber\\
&\left.-2\frac{\mbf{k}^2\mbf{l}^2-2(\mbf{k}\cdot \mbf{l})^2}{(\mbf{k}^2+m_{\pi}^2)(\mbf{l}^2+m_{\pi}^2)}\left(\mathcal{T}_1-\frac{\mathcal{A}_{\pi}(\mathbf{k}^{2})}{\mbf{k}^2+m_{\pi}^2}\right)\left(\mathcal{T}_2-\frac{\mathcal{A}_{\pi}(\mathbf{l}^{2})}{\mbf{l}^2+m_{\pi}^2}\right)\right].
\end{flalign}

In order to simplify the 15-dimensional phase space integration in Eq.~\ref{emiss}, we go to the centre of mass frame by making the following coordinate transformations,
\begin{align}
    \mbf{p_1}=\mbf{p}+\mbf{p_i}, ~~~~~~ \mbf{p_2}=\mbf{p}-\mbf{p_i}, ~~~~~ \mbf{p_3}=\mbf{p}+\mbf{p_f}, ~~~~~~ \mbf{p_4}=\mbf{p}-\mbf{p_f}.
\end{align}
Here $ \mbf{p} = \mbf{p}_{1} + \mbf{p}_{2} \approx \mbf{p}_{3} + \mbf{p}_{4}$ in the limit $\mbf{p}_{i} \gg \mbf{k}_{s}$. In this new frame we can further simplify the pion propagator expansion in Eq.~\ref{prop appro 2}. It can be shown that $k_{0} = \omega/2+\mathbf{p}.\mathbf{k}/m_{N}$ giving,
\begin{equation}
    2k\cdot k_s=\omega^2+2\mbf{p}\cdot \mbf{k}\frac{\omega}{m_N}-2\mbf{k}\cdot \mbf{k_s}.
\end{equation}
Neglecting the subdominant term $2\mbf{p}\cdot \mbf{k}\frac{\omega}{m_N}$, which is suppressed by the factor $m_{N}$, the pion propagator in Eq.~\ref{prop appro 2} can be written as,
\begin{align}
    \frac{1}{(k-k_s)^2-m_{\pi}^2} &\approx \frac{1}{k^2-m_{\pi}^2}\left(1-\frac{m_s^2-\omega^2+2\mbf{k}\cdot \mbf{k_s}}{k^2-m_{\pi}^2}\right) \nonumber \\
    & \approx \frac{1}{k^2-m_{\pi}^2}\left(1-\frac{-\mbf{k_s}^2+2\mbf{k}\cdot \mbf{k_s}}{k^2-m_{\pi}^2}\right).
\end{align}
$\mathcal{R}$ in Eq.~\ref{simplifying terms} can also be simplified as,
\begin{equation}
    \mathcal{R} \approx -2(\mbf{p_f}-\mbf{p_i})\cdot\mbf{k_s}~(\mbf{p_f}+\mbf{p_i})\cdot\mbf{k_s}.
\end{equation}

Using the following redefinition,
\begin{align}
    u\equiv \frac{\mbf{p_i}^2}{m_N T}, \quad v\equiv\frac{\mbf{p_f}^2}{m_N T}, \quad x\equiv\frac{\omega}{T}, \quad y\equiv\frac{m_{\pi}^2}{m_N T}, \quad q\equiv\frac{m_s}{T}, \quad z\equiv\cos\theta_{if},
\end{align}
we perform the phase space integral similar to Ref.~\cite{Dent:2012mx,Giannotti:2005tn}. This gives the instantaneous luminosity per unit volume as,
\begin{align}
    \frac{dL}{dV}=&~\frac{n_B^2 T^{7/2}}{2^{14} \pi^{13/2}m_N^{5/2}}\int_{q/\eta}^{\infty} du\int_{0}^{\infty} dv\int_{q/\eta}^{\infty} dx\int S\sum_{spins}|\mathcal{M}|^2 \nonumber\\
   &\times x\sqrt{uv}\sqrt{x^2-q^2}e^{-u}\delta(u-v-x) d\Omega_id\Omega_fd\Omega_s,
\end{align}
and MFP due to absorption as,
\begin{align}
\lambda^{-1}_{abs}=&~\frac{n_B^2 }{2^{9} \pi^{7/2}m_N^{5/2}T^{1/2}\sqrt{x^2-q^2}}\int_{0}^{\infty} du\int_{q/\eta}^{\infty} dv\int S\sum_{spins}|\mathcal{M}(-k_s)|^2 \nonumber \\
&\times \sqrt{uv}e^{-u}\delta(u-v+x) d\Omega_id\Omega_f.
\end{align}
While deriving the above equation, we have taken advantage of the fact that our amplitude doesn't explicitly depend explicitly on $\mbf{p}$. All the angular integration except the $z$ can be done analytically. We do not show the full angular integration as the expression for the amplitude is quite large. Instead, we summarize the relevant integrals as follows, 
\begin{equation}
    I_{a,b,c}=\int d\Omega_id \Omega_fd\Omega_s(\mbf{\hat{p_i}}\cdot\mbf{\hat{p_f}})^a(\mbf{\hat{p_i}}\cdot\mbf{\hat{k_s}})^b(\mbf{\hat{p_f}}\cdot\mbf{\hat{k_s}})^c=\int_{-1}^1f_{a,b,c}(z)dz,
\end{equation}
where the function $f_{a,b,c}(z)$ is given in Tab.~\ref{ang_int}.

\begin{table}[t]
\[
\scalebox{0.75}{
$f_{a,b,c}(z)=
\begin{array}{|c||c|c|c|c|c|}
\hline
 & a = 0 & a = 1 & a = 2 & a = 3 & a = 4 \\
\hline\hline
b = 0 &
\begin{array}{c}
32\pi^3 \\
\frac{32\pi^3}{3} \\
\frac{32\pi^3}{5}
\end{array} &
\begin{array}{c}
\frac{32\pi^3 z}{3} \\
\frac{32\pi^3 z}{5}
\end{array} &
\begin{array}{c}
8\pi^2 \left(\frac{4\pi z^2}{3} + \frac{4\pi(1 - z^2)}{3} \right) \\
8\pi^2 \left(\frac{4\pi z^2}{5} + \frac{4\pi(1 - z^2)}{15} \right) \\
8\pi^2 \left(\frac{4\pi z^2}{7} + \frac{4\pi(1 - z^2)}{35} \right)
\end{array} &
\begin{array}{c}
8\pi^2 \left(\frac{4\pi z^3}{5} + \frac{4\pi z(1 - z^2)}{5} \right) \\
8\pi^2 \left(\frac{4\pi z^3}{7} + \frac{12\pi z(1 - z^2)}{35} \right)
\end{array} &
\begin{array}{c}
8\pi^2 \left(\frac{4\pi z^4}{5} + \frac{8\pi z^2(1 - z^2)}{5} + \frac{4\pi (1 - z^2)^2}{5} \right) \\
8\pi^2 \left(\frac{4\pi z^4}{7} + \frac{24\pi z^2(1 - z^2)}{35} + \frac{4\pi (1 - z^2)^2}{35} \right) \\
8\pi^2 \left(\frac{4\pi z^4}{9} + \frac{8\pi z^2(1 - z^2)}{21} + \frac{4\pi (1 - z^2)^2}{105} \right)
\end{array} \\
\hline
b = 1 &
\begin{array}{c}
32\pi^3 z \\
\frac{32\pi^3 z}{3} \\
\frac{32\pi^3 z}{5}
\end{array} &
\begin{array}{c}
\frac{32\pi^3 z^2}{3} \\
\frac{32\pi^3 z^2}{5}
\end{array} &
\begin{array}{c}
8\pi^2 \left(\frac{4\pi z^3}{3} + \frac{4\pi z(1 - z^2)}{3} \right) \\
8\pi^2 \left(\frac{4\pi z^3}{5} + \frac{4\pi z(1 - z^2)}{15} \right) \\
8\pi^2 \left(\frac{4\pi z^3}{7} + \frac{4\pi z(1 - z^2)}{35} \right)
\end{array} &
\begin{array}{c}
8\pi^2 \left(\frac{4\pi z^4}{5} + \frac{4\pi z^2(1 - z^2)}{5} \right) \\
8\pi^2 \left(\frac{4\pi z^4}{7} + \frac{12\pi z^2(1 - z^2)}{35} \right)
\end{array} &
\begin{array}{c}
8\pi^2 \left(\frac{4\pi z^5}{5} + \frac{8\pi z^3(1 - z^2)}{5} + \frac{4\pi z(1 - z^2)^2}{5} \right) \\
8\pi^2 \left(\frac{4\pi z^5}{7} + \frac{24\pi z^3(1 - z^2)}{35} + \frac{4\pi z(1 - z^2)^2}{35} \right) \\
8\pi^2 \left(\frac{4\pi z^5}{9} + \frac{8\pi z^3(1 - z^2)}{21} + \frac{4\pi z(1 - z^2)^2}{105} \right)
\end{array} \\
\hline
b = 2 &
\begin{array}{c}
32\pi^3 z^2 \\
\frac{32\pi^3 z^2}{3} \\
\frac{32\pi^3 z^2}{5}
\end{array} &
\begin{array}{c}
\frac{32\pi^3 z^3}{3} \\
\frac{32\pi^3 z^3}{5}
\end{array} &
\begin{array}{c}
8\pi^2 \left(\frac{4\pi z^4}{3} + \frac{4\pi z^2(1 - z^2)}{3} \right) \\
8\pi^2 \left(\frac{4\pi z^4}{5} + \frac{4\pi z^2(1 - z^2)}{15} \right) \\
8\pi^2 \left(\frac{4\pi z^4}{7} + \frac{4\pi z^2(1 - z^2)}{35} \right)
\end{array} &
\begin{array}{c}
8\pi^2 \left(\frac{4\pi z^5}{5} + \frac{4\pi z^3(1 - z^2)}{5} \right) \\
8\pi^2 \left(\frac{4\pi z^5}{7} + \frac{12\pi z^3(1 - z^2)}{35} \right)
\end{array} &
\begin{array}{c}
8\pi^2 \left(\frac{4\pi z^6}{5} + \frac{8\pi z^4(1 - z^2)}{5} + \frac{4\pi z^2(1 - z^2)^2}{5} \right) \\
8\pi^2 \left(\frac{4\pi z^6}{7} + \frac{24\pi z^4(1 - z^2)}{35} + \frac{4\pi z^2(1 - z^2)^2}{35} \right) \\
8\pi^2 \left(\frac{4\pi z^6}{9} + \frac{8\pi z^4(1 - z^2)}{21} + \frac{4\pi z^2(1 - z^2)^2}{105} \right)
\end{array} \\
\hline
b = 3 &
\begin{array}{c}
32\pi^3 z^3 \\
\frac{32\pi^3 z^3}{3} \\
\frac{32\pi^3 z^3}{5}
\end{array} &
\begin{array}{c}
\frac{32\pi^3 z^4}{3} \\
\frac{32\pi^3 z^4}{5}
\end{array} &
\begin{array}{c}
8\pi^2 \left(\frac{4\pi z^5}{3} + \frac{4\pi z^3(1 - z^2)}{3} \right) \\
8\pi^2 \left(\frac{4\pi z^5}{5} + \frac{4\pi z^3(1 - z^2)}{15} \right) \\
8\pi^2 \left(\frac{4\pi z^5}{7} + \frac{4\pi z^3(1 - z^2)}{35} \right)
\end{array} &
\begin{array}{c}
8\pi^2 \left(\frac{4\pi z^6}{5} + \frac{4\pi z^4(1 - z^2)}{5} \right) \\
8\pi^2 \left(\frac{4\pi z^6}{7} + \frac{12\pi z^4(1 - z^2)}{35} \right)
\end{array} &
\begin{array}{c}
8\pi^2 \left(\frac{4\pi z^7}{5} + \frac{8\pi z^5(1 - z^2)}{5} + \frac{4\pi z^3(1 - z^2)^2}{5} \right) \\
8\pi^2 \left(\frac{4\pi z^7}{7} + \frac{24\pi z^5(1 - z^2)}{35} + \frac{4\pi z^3(1 - z^2)^2}{35} \right) \\
8\pi^2 \left(\frac{4\pi z^7}{9} + \frac{8\pi z^5(1 - z^2)}{21} + \frac{4\pi z^3(1 - z^2)^2}{105} \right)
\end{array} \\
\hline
b = 4 &
\begin{array}{c}
32\pi^3 z^4 \\
\frac{32\pi^3 z^4}{3} \\
\frac{32\pi^3 z^4}{5}
\end{array} &
\begin{array}{c}
\frac{32\pi^3 z^5}{3} \\
\frac{32\pi^3 z^5}{5}
\end{array} &
\begin{array}{c}
8\pi^2 \left(\frac{4\pi z^6}{3} + \frac{4\pi z^4(1 - z^2)}{3} \right) \\
8\pi^2 \left(\frac{4\pi z^6}{5} + \frac{4\pi z^4(1 - z^2)}{15} \right) \\
8\pi^2 \left(\frac{4\pi z^6}{7} + \frac{4\pi z^4(1 - z^2)}{35} \right)
\end{array} &
\begin{array}{c}
8\pi^2 \left(\frac{4\pi z^7}{5} + \frac{4\pi z^5(1 - z^2)}{5} \right) \\
8\pi^2 \left(\frac{4\pi z^7}{7} + \frac{12\pi z^5(1 - z^2)}{35} \right)
\end{array} &
\begin{array}{c}
8\pi^2 \left(\frac{4\pi z^8}{5} + \frac{8\pi z^6(1 - z^2)}{5} + \frac{4\pi z^4(1 - z^2)^2}{5} \right) \\
8\pi^2 \left(\frac{4\pi z^8}{7} + \frac{24\pi z^6(1 - z^2)}{35} + \frac{4\pi z^4(1 - z^2)^2}{35} \right) \\
8\pi^2 \left(\frac{4\pi z^8}{9} + \frac{8\pi z^6(1 - z^2)}{21} + \frac{4\pi z^4(1 - z^2)^2}{105} \right)
\end{array} \\
\hline
\end{array}
$
}
\]
\caption{Angular integration table. Boxes with three entries correspond to $c=0,2,4$, while the boxes with two entries correspond to $c=1,3$.}
\label{ang_int}
\end{table}

\bibliographystyle{JHEP}
\bibliography{snhiggs}

\end{document}